\documentclass[prl,superscriptaddress,twocolumn,footinbib]{revtex4-2}
\pdfoutput=1
\usepackage[colorlinks=true,urlcolor=blue]{hyperref}
\usepackage{amsfonts}
\usepackage{graphicx}
\usepackage{placeins}
\usepackage{amsmath}
\usepackage{xcolor}
\usepackage{color}
\usepackage{bm}
\usepackage{ulem}
\usepackage{array}
\usepackage{epsfig}
\usepackage{dutchcal}
\usepackage{soul}
\usepackage{multirow}

\definecolor{red}{rgb}{0.75,0,0}
\definecolor{blue}{rgb}{0,0,0.75}
\definecolor{green}{rgb}{0,0.5,0}

\begin{document}
	
	\title{Passive defect driven morphogenesis in nematic membranes}
	
	\author{D. J. G. Pearce}
     \email{daniel.pearce@unige.ch}
	\affiliation{Dept.~of Theoretical Physics, University of Geneva, 1211 Geneva, Switzerland}

\author{C. Thibault}
	\affiliation{Laboratoire Physico-Chimie Curie, Institut Curie, Universit\' e PSL, Sorbonne Universit\' e, CNRS UMR168, F-75248 Paris, France}
 
 \author{Q. Chaboche}
	\affiliation{Laboratoire Physico-Chimie Curie, Institut Curie, Universit\' e PSL, Sorbonne Universit\' e, CNRS UMR168, F-75248 Paris, France}
 	
	\author{C. Blanch-Mercader}
     \email{carles.blanch-mercader@curie.fr}
	\affiliation{Laboratoire Physico-Chimie Curie, Institut Curie, Universit\' e PSL, Sorbonne Universit\' e, CNRS UMR168, F-75248 Paris, France}
    
\begin{abstract}
    Orientational order is a common feature of many biological and synthetic materials. Topological defects are discontinuities in this order that are often coupled to geometric features of the materials. Here, we study the equilibrium shapes of fluid membranes featuring a $+1$ topological defect as a model for morphogenesis. We show, through simulation and analytic calculation, that the membrane can spontaneously deform toward a conical shape with a defect at its apex. We show that the relative stability of the deformation is controlled by the balance of the elastic parameters. When boundary constraints are introduced, we observe three distinct modes of deformation. These deformation modes take advantage of the way in which splay, twist and bend distortions of the director field can be exchanged on a curved surface. Finally we demonstrate inverted solutions. Our findings demonstrate a mechanism for passive defect driven morphogenesis as well as the fusion of $+1/2$ topological defect pairs on deformable surfaces.
\end{abstract}

	\maketitle


    The study of shape shifting materials is often inspired by biological systems and has a broad range of applications. For example, deformable solids that have a prescribed orientational field, such as elastomers \cite{buguin2006micro,ware2015voxelated,aharoni2018universal} or inflatable structures \cite{pikul2017stretchable,siefert2019bio}, can achieve families of morphologies by differential growth \cite{dervaux2008morphogenesis,muller2008conical,modes2010disclination,modes2011blueprinting,siefert2020programming} and have applications in soft robotics\cite{lee2017soft}. Here, we focus on fluid membranes with nematic order. Unlike the previous cases, these membranes can, in addition, flow, remodel and adapt both their orientational field and shape.   

    Fluid membranes achieve their shapes by minimizing a free-energy subjected to physical constraints \cite{zhong1989bending,lipowsky1991conformation,julicher1994shape,guven2018geometry}. When orientational order is included, the stresses generated by topological defects can render a flat surface unstable to out-of-plane perturbations \cite{frank2008defects}. In addition, the geometrical properties of the surface, such as the extrinsic curvature, can in turn influence the dynamics of the orientational field and its equilibrium configurations \cite{napoli2012surface,napoli2012extrinsic,pearce2022coupling,segatti2014equilibrium,jesenek2015defect,nitschke2018nematic,nestler2018orientational,nestler2020properties}. Indeed, for prescribed geometries, couplings between the nematic field and the intrinsic or the extrinsic geometry have been shown to induce symmetry-breaking of orientational configurations, or influence topological defect dynamics, \cite{napoli2012surface,napoli2012extrinsic,vafa2022defect,santiago2018stresses,santiago2019membrane,pearce2022coupling}. Recent studies extended these works by, for instance, including out-of-equilibrium processes, such as active stresses or active torques, or varying surface geometry and topology \cite{pearce2019geometrical,pearce2020defect,khoromskaia2023active,nestler2021active,hoffmann2022theory,salbreux2022theory,bell2022active,vafa2023periodic,singha2023clustering}.  

\begin{figure}[t]
        \centering
        \includegraphics[width=\columnwidth]{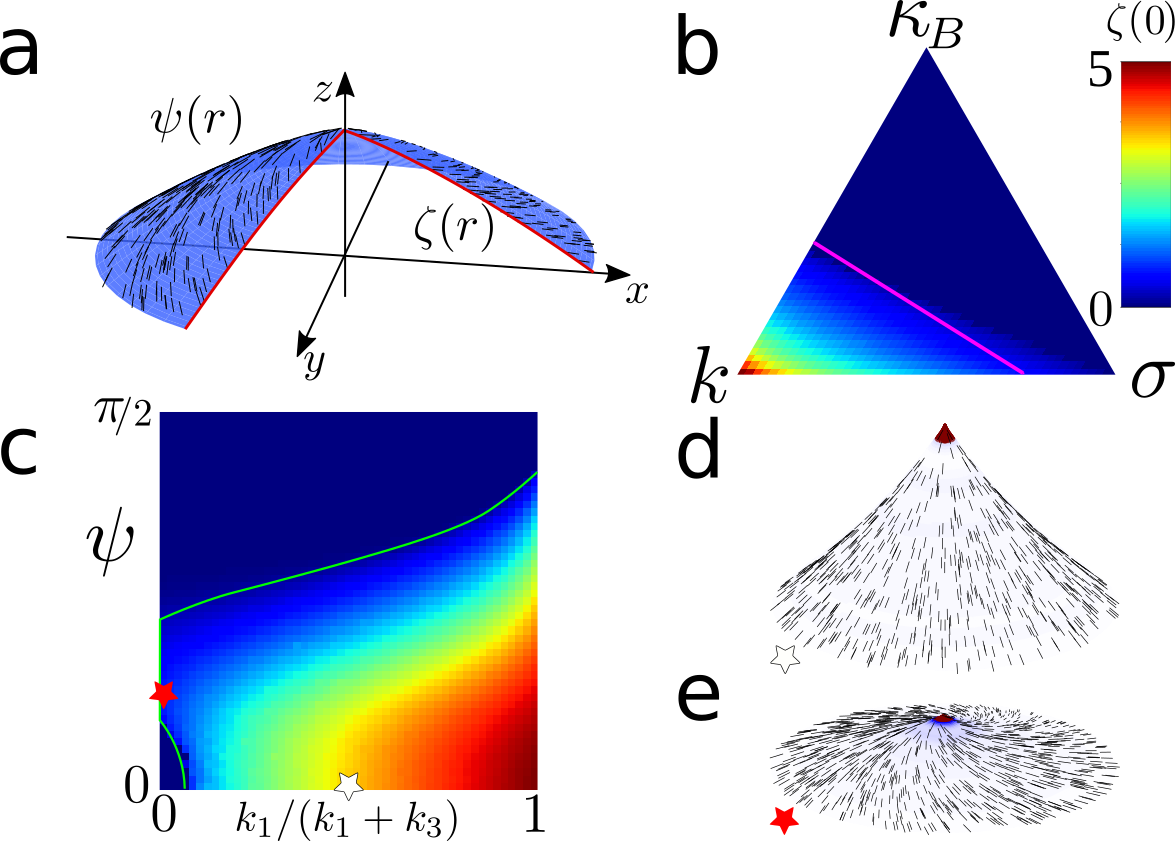}
        \caption{\label{fig:f1} (a) Geometrical setup. The red line shows the height profile, $\zeta(r)$, and black dashes show the orientation of the director, $\psi(r)$. (b) The height of a membrane featuring a single aster defect with equal Frank constants $k_1\!=\!k_2\!=\!k_3\!=\!k/3$ as a function of $k$, surface tension $\sigma$ and bending rigidity $k_B$ with the constraint $\sigma+k_B+k=1$. The magenta line corresponds to Eq.~\ref{eq:5}. From this point forward we fix $\sigma=k_B=1/10$ and $k_1+k_2+k_3=1$ in all results unless otherwise stated. (c) Height of a membrane as a function of the prescribed uniform phase, $\psi$, and ratio between the splay and bend coefficients, ($k_2=0$). The green line corresponds to the predicted transition to conical surfaces, see SI. (d)\&(e) Examples of equilibrium surfaces with parameters given by: white star ($k_1=k_3$, $\psi=0$), red star ($k_3=1$, $\psi=\pi/8$). (d) An aster on a conical surface, and (e) a spiral on a concave surface. Blue (red) regions: negative (positive) Gaussian curvature.}
    \end{figure}

    We study rotationally symmetric systems, which therefore feature a $+1$ topological defect at their center; $+1$ topological defects have been associated with geometrical changes in natural and synthetic systems~\cite{maroudas2021topological,keber2014topology,blanch2021quantifying,ravichandran2024topology}. We use cylindrical coordinates to define these shapes, where $r$ is the radial coordinate, $\theta$ is the azimuthal coordinate and $\zeta$ is the axial coordinate, Fig.~\ref{fig:f1}a. The outer circular boundary is placed at $r=R$; without loss of generality, we set the vertical offset by $\zeta(R)=0$ and $R=1$. The orientation of the nematic field is described by a director field $\hat{n}$ that represents the averaged local orientation on the surface. We consider that the director field is tangential to the surface. In addition, we consider that the system is deep into the nematic phase and impose $|\hat{n}|=1$. This allows the director field to be defined by a scalar phase $\psi(r)$, which corresponds to the angle between the director field and the curvilinear radial direction. The cases $\psi=0$ corresponds to the aster, $0<\psi<\pi/2$ to a spiral, and $\psi=\pi/2$ to the vortex.

    The two-dimensional free-energy of a fluid membrane with nematic order is given by
    \begin{eqnarray}
	\mathcal{F}&=&\int_\mathcal{A}\Big\{k_B{\cal H}^2+\sigma+k_1(\nabla\cdot \hat{n})^2\nonumber \\
&+&k_2(\hat{n}\cdot(\nabla\times \hat{n}))^2+k_3(\hat{n}\times(\nabla\times \hat{n}))^2\Big\}\textrm{d}a. \label{eq:2}
    \end{eqnarray}
    The first term is the bending energy with mean curvature ${\cal H}$, and the second term represents surface tension. We disregard anisotropies in bending energy dependent on $\hat{n}$. The other terms are the Frank free-energy associated respectively with splay, twist and bend distortions of the director field $\hat{n}$ \cite{frank1958liquid,de1993physics}. The corresponding elastic coefficients are: the bending rigidity $k_B$, the surface tension $\sigma$, and the reduced Frank constants $k_1$, $k_2$, and $k_3$ that are proportional to the membrane thickness. Because the director field is parallel to the surface, the effects of the saddle-splay distortions with elastic constant $k_{24}$ can be absorbed in a redefinition of the other Frank constants, \cite{napoli2012extrinsic}. As shown in Refs.~\cite{napoli2012surface,napoli2012extrinsic} and subsequently extended for more general cases in \cite{nestler2019finite}, the thin-film limit of the Frank free-energy results in contributions that couple the director field with both the intrinsic and the extrinsic geometry. In our case, Eq.~\eqref{eq:2} takes the form derived in Refs.~\cite{napoli2012surface,napoli2012extrinsic} and the expression of the free-energy \eqref{eq:2} for the special case of a surface of revolution is derived in Sec.~1.1 \cite{supp_mater}. Other descriptions for fluid surfaces with orientational order 
    neglected the couplings with the extrinsic geometry \cite{frank2008defects,santiago2018stresses,vafa2022defect}. For a discussion on the thin-film approximation of liquid crystals, we refer to Ref.~\cite{nestler2019finite}.

    We combine analytical and numerical approaches to study equilibrium configurations of the free-energy \eqref{eq:2}. The former approach is restricted to director fields with a uniform phase $\psi$, see Sec. 1 and 2 in \cite{supp_mater}. The latter approach discretizes the functions $\zeta(r)$ and $\psi(r)$, Fig.~\ref{fig:f1}a, and uses a Monte-Carlo algorithm, see Sec. 3 \cite{supp_mater}. 

    We first consider the case where all the reduced Frank constants are equal, i.e. the one-constant approximation $k_1\!=\!k_2\!=\!k_3\!=\!k/3$. The energy scale is set by the constraint $k\!+\!k_B\!+\!\sigma\!=\!1$. Fig.~\ref{fig:f1}b shows the height at the center $\zeta(0)$ of the equilibrium states that were found for varying elastic parameters. When $k_B$ or  $\sigma$ dominate, the minimal states are flat discs which minimize surface area and curvature. Because the Frank constants are equal, $k_1\!=\!k_3$, all director configurations with a constant phase $\psi$ have equal energy, and thus are minimal states, \cite{de1993physics}. However, when $k$ dominates, the minimal state changes to an approximately conical surface with phase $\psi=0$, Fig.~\ref{fig:f1}d. The morphological transition occurs via a symmetry-breaking process during which the director adopts an aster configuration. 

    To better understand this spontaneous transition from flat to conical surfaces, we derived analytically the normal force balance equation for a surface of revolution with an embedded director field with uniform phase $\psi$, see Sec.~1.2 in \cite{supp_mater} (see \cite{santiago2019membrane} for a general case). In the special case $\sigma=0$, this nonlinear ODE has a set of exact non-trivial solutions corresponding to conical surfaces of varying heights; this is valid even when the one-constant approximation is relaxed, see Sec. 1.3 in \cite{supp_mater}. This gives us a subset of shapes described by $\zeta(r)=\pm m (r-1)$, with $m$ the vertical distance of the tip of the cone from the base, that allow us to move continuously from a flat disk to a cone of varying height. Minimizing Eq.~\eqref{eq:2} with respect to the height $m$ and for a fix phase $\psi$, one obtains the non-trivial condition 
    \begin{equation}
  	m^2=\frac{A(1-2\sin(\psi)^2)-2-B}{1+A\sin{(\psi)}^2+ B},\label{eq:3}
    \end{equation}
    where $A=4 k/3k_B$ and $B=(4\sigma/k_B)(1-\Delta^2)/\log(1/\Delta)$ are dimensionless parameters and $\Delta$ is a dimensionless cut-off lengthscale. The logarithmic divergence at $\Delta=0$ is commonly found in the context of topological defects \cite{de1993physics} and also describes the divergence in curvature at the tip of the conical surface. We explore the effect of $\Delta$ on our results in Sec.~2 in \cite{supp_mater}. Eq.~\eqref{eq:3} leads to the existence condition for conical shapes $A(1-2\sin(\psi)^2)>2+B$. As observed, Eq.~\eqref{eq:3} shows that as $k/k_B$ or $k/\sigma$ increase, the minimal surface varies from a flat disc ($m=0$) to a cylinder ($m\rightarrow\infty$). The total free-energy associated with this minimal surface reads
    \begin{equation}
        \frac{{\cal F}_c}{\pi k_B }=\sqrt{(1+A\sin{(\psi)}^2+B)(A\cos{(\psi)}^2-1)}\log(1/\Delta).\label{eq:4}
    \end{equation}
    In addition, Eq.~\eqref{eq:4} shows that the minimal phase corresponds to $\psi=0$, even when the Frank constants $k_1\!=\!k_3$ are equal. The selection mechanism for the phase arises from the coupling between the director field and the extrinsic curvature, which tend to align the director field with the minimal principal curvature \cite{napoli2012extrinsic}. In the case of conical surfaces, the aster is favoured because it features only splay distortions, and vanishing twist and bend distortions. Furthermore, the total Frank free-energy for the aster decreases as the height of the cone increases, leading to the spontaneous out-of-plane deformation of a surface. This is balanced by the increased curvature and area of the surface. The threshold for a flat disc with an aster topological defect to become unstable is set when its energy (i.e. $2\pi k/3 \log(1/\Delta)+\sigma \pi(1-\Delta^2)$) equals the energy \eqref{eq:4} for $\psi=0$, that is when
    \begin{equation}
        k/3 = \frac{k_B}{2}+\sigma\frac{(1-\Delta^2)}{\log(1/\Delta),}\label{eq:5}
    \end{equation}
    shown with the magenta line, Fig.~\ref{fig:f1}b. This condition equals the existence condition of conical shapes for $\sigma=0$, Sec.~1.3 in \cite{supp_mater}. A similar approach is used with varying elastic coefficients and defect phase to generate the green line in Fig.~\ref{fig:f1}c.
    
    According to Eq.~\ref{eq:3}, the height of the surface is also controlled by the phase of the defect. For equal Frank constants, the existence condition for conical shapes can be satisfied when $\psi<\pi/4$. Beyond this point, the increased energy associated with twist and bend distortions of the director field on a conical surface is too great and a flat surface is favoured. To further explore this effect, we numerically studied the equilibrium shapes of surfaces with a prescribed uniform phase, $\psi$, and a varying ratio of bend and splay elastic constants, $k_1/(k_1+k_3)$, Fig.~\ref{fig:f1}c. In general, we find that vortices cannot deform a surface. In most cases, the aster generates the maximal out-of-plane deformation, except when $k_3\gg k_1$ and the maximal deformation occurs near $\psi\sim\pi/8$, see Fig.~\ref{fig:f1}e. This result is also predicted by the conical surface approximation, which is given by the green line in Fig.~\ref{fig:f1}c.
		
\begin{figure}[t]
	\centering
	\includegraphics[width=\columnwidth]{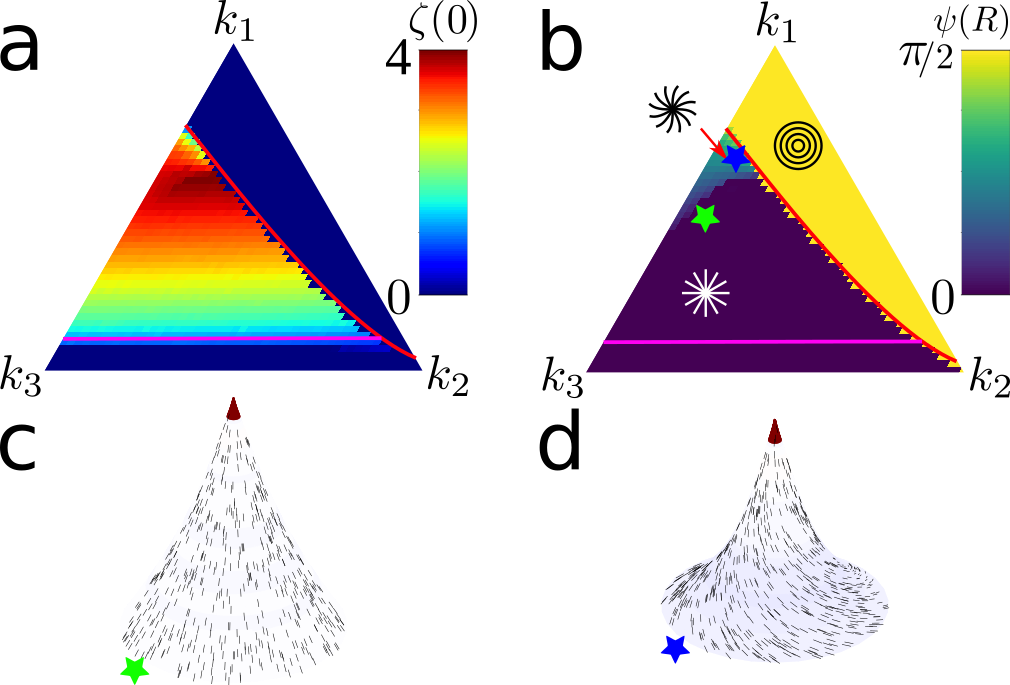}
	\caption{\label{fig:f2} (a) Height at the center and (b) phase at the boundary of the surface as a function of the three Frank constants with the constraint $k_1+k_2+k_3=1$. Magenta line is Eq.~\eqref{eq:5} and red line is Eq.~\eqref{eq:6}. (c)\&(d) Example surfaces with parameters: green star ($k_1=1/2$, $k_2=1/10$, $k_3=4/10$) blue star ($k_1=13/20$, $k_2=1/10$, $k_3=1/4$). (c) shows a conical aster surface while (d) features a concave surface with an aster at the center and spiral at the boundary. Blue (red) regions: negative (positive) Gaussian curvature. In all panels, $k_B=\sigma=1/10$.}
\end{figure}

    To further explore these results, we now focus on the effects of varying all three Frank elastic constants; we also relax the constraint that $\psi$ is constant. Both $\zeta'$ and $\psi$ are only constrained by the rotational symmetry at the boundary. Figs.~\ref{fig:f2}a and \ref{fig:f2}b show respectively the height at the center, $\zeta(0)$, and the phase at the outer boundary, $\psi(R)$, of the energy minimizing states for varying Frank coefficients. We found that the height of the deformed membrane is determined by the magnitude of $k_1$. Two transitions from flat to deformed surfaces were identified, which are primarily dependent on the relative values of $k_1$ and $k_3$. The transition from a flat to a deformed surface with an aster is determined by the threshold \eqref{eq:5} with $k_1=k/3$ (magenta line in Figs.~\ref{fig:f2}a-b). A flat disc with a vortex is linearly stable to out-of-plane deformations, see Sec.~1.4 in \cite{supp_mater}. The red line in Fig.~\ref{fig:f2}a-b can be found by comparing the energy of a flat disc with a vortex (i.e. $2\pi k_3 \log(1/\Delta)+\pi \sigma(1-\Delta^2)$) to the energy \eqref{eq:4} for $\psi=0$, that is when
    \begin{equation}
           \frac{k_1}{k_B} = \frac{(k_3/k_B+B/4)^2}{(1+B)} + \frac{1}{4}. \label{eq:6}
    \end{equation}

    On a flat surface, if $k_1\!>\!k_3$ ($k_1\!<\!k_3$), the director field assumes a vortex (aster) configuration \cite{de1993physics}. On a curved surface, however, an aster can be energetically favoured for values of $k_1>k_3$; indeed all deformed surfaces here feature an aster at their core. In most cases, this is a conical aster deformation with constant phase, Fig.~\ref{fig:f2}c. However, when $k_2\ll k_3<k_1$ we observe a new state with a spatially varying phase which features a conical aster close to the core combined with a negative Gaussian curvature spiral region near the boundary, Fig.~\ref{fig:f2}d. The spiral region reduces splay at the cost of additional bend, however the bend is then exchanged for twist on the negative Gaussian curvature surface when $\psi$ has an intermediate value, which significantly reduces the free-energy density when $k_2$ is small. 

\begin{figure}[t]
	\centering
	\includegraphics[width=\columnwidth]{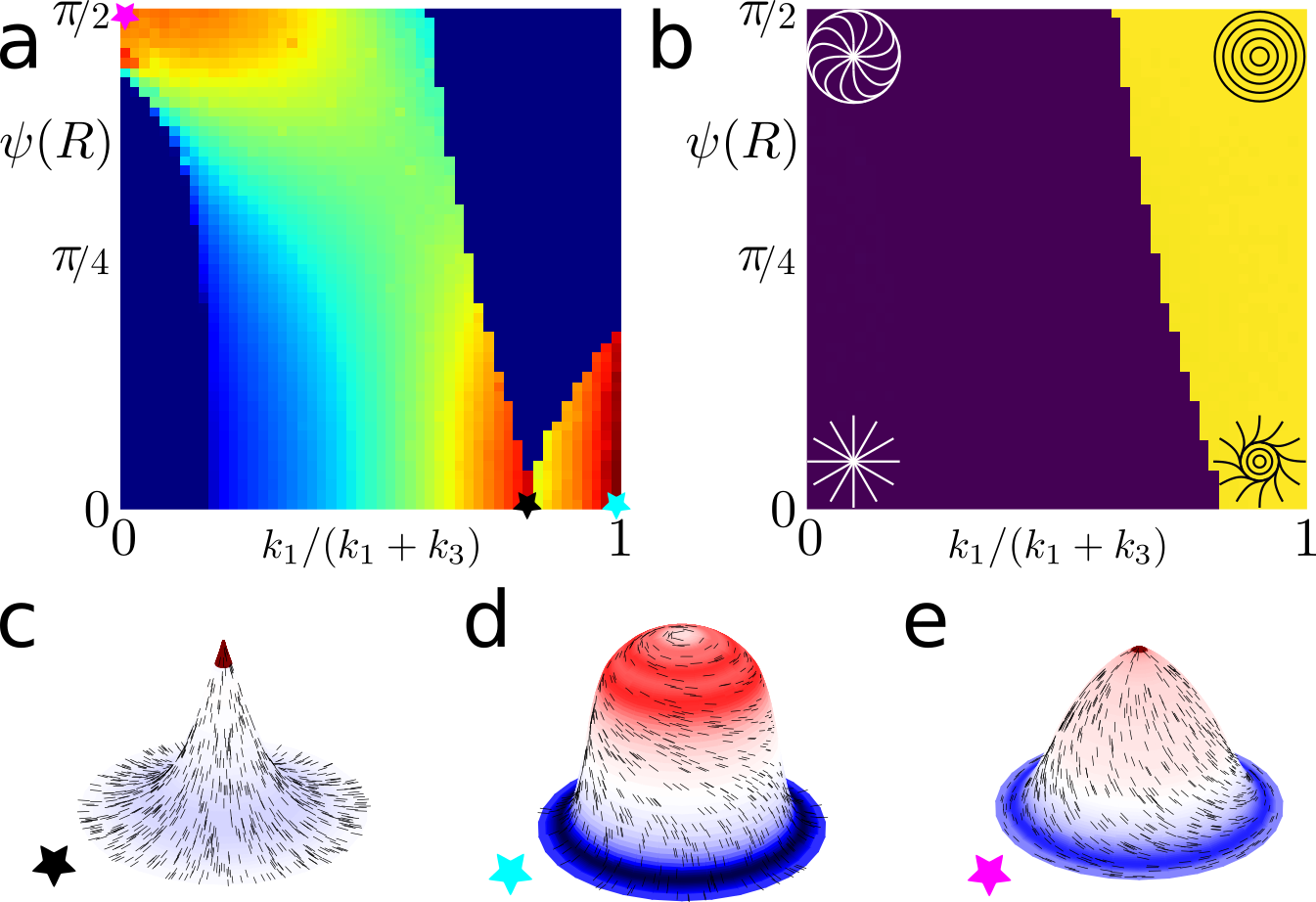}
	\caption{\label{fig:f3}(a) Height and (b) phase of the director field at the center of the membrane as a function of the phase at the boundary, $\psi(R)=\psi_R$, and the ratio between bend and splay coefficients ($k_2=1/3$) for surfaces with fix boundaries $\zeta'(R)=0$. (c-e) Example of the three modes of out-of-plane deformation with fix boundary conditions: (c) pointy with an aster, (d) domed with an aster-to-vortex transition and (e) pointy with a vortex-to-aster transition. The states in (c-e) correspond to the black ($k_1 = 1/2$, $k_3 = 1/6$, $\psi_R=0$), turquoise ($k_1 = 2/3$, $k_3 = 0$, $\psi_R=0$) and pink ($k_1 = 0$, $k_3 = 2/3$, $\psi_R=\pi/2$) stars in (a) respectively. Blue (red) regions: negative (positive) Gaussian curvature. In all panels, $k_B=\sigma=1/10$.}
\end{figure}

    Now, we consider the system under boundary conditions that might be found in biological systems. Hence, we fix $\psi(R)\!=\!\psi_R$ as the phase at the boundary of the membrane and fix $\zeta'(R)\!=\!0$, implying that the surface must be flat at its boundary \footnote{Note that the boundary condition $\zeta'(R)\!=\!0$ enforces that the total Gaussian curvature, including the tip, is zero.}. Figs.~\ref{fig:f3}a\&b show the height and the phase at the center of the surface for a full range of the bend and splay constants ($k_2=1/3$) and the boundary phase $\psi_R$. On a flat surface, the value of $k_1/(k_1+k_3)$ controls the phase at the core of the topological defect. By varying $\psi_R$, we can explore regimes where the preferred phase at the center is frustrated with the boundary. We observe five distinct states. First, when the boundary phase $\psi_R$ is compatible with the dominant elastic coefficient, we observe flat asters and flat vortices. In addition to this there are three deformed configurations.     
	
    When $k_1\!\approx\!k_3$ and $\psi_R\!\approx\!0$ we find pointy surfaces featuring an aster, where the core and boundary of the defect are broadly in phase, see Fig.~\ref{fig:f3}c. As in Fig.~\ref{fig:f2}, the deformation here reduces splay distortions and hence stabilizes an aster, even in some cases when $k_1>k_3$. 

    \begin{figure}[t]
        \centering
        \includegraphics[width=\columnwidth]{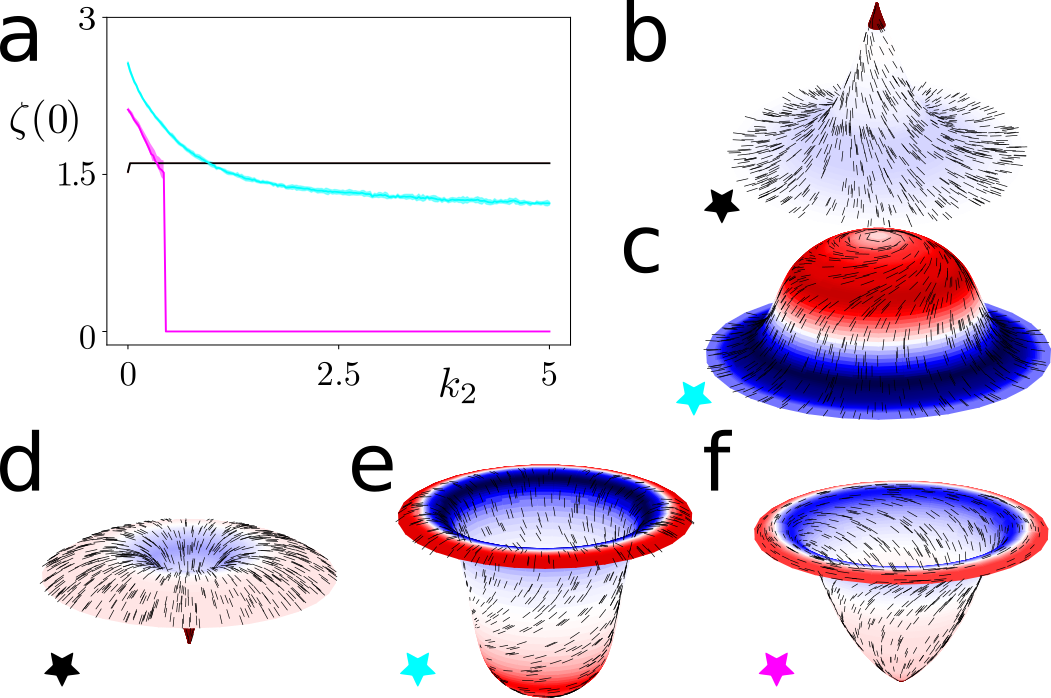}
        \caption{\label{fig:f4} (a) Height at the apex as a function of the magnitude of $k_2$ for the three modes of out-of-plane deformation in Fig.~\ref{fig:f3}c-e. The values of $k_1$, $k_3$ and $\psi_R$ are as outlined previously in caption Fig.~\ref{fig:f3}. Here the constraint $k_1+k_2+k_3=1$ has been relaxed to allow $k_2$ to vary. (d-e) Examples of inverted versions for the three main deformation modes with $\zeta'(R)=1$. Blue (red) regions: negative (positive) Gaussian curvature. In all panels, $k_B=\sigma=1/10$.}
    \end{figure}

    In the bend dominated regime $k_1\!<\!k_3$ and for $\psi_R\!\approx\!\pi/2$, we observe pointy deformations on which the phase transitions from a vortex at the boundary to an aster at the center, Fig.~\ref{fig:f3}e. In this configuration, the transition in the phase is localized to a ring of negative Gaussian curvature where bend distortions can be reduced at the cost of twist through the previously highlighted mechanism highlighted in Fig.~\ref{fig:f2}d.
    
    In the splay dominated regime $k_1\!>\!k_3$ and for $\psi_R\!\approx\!0$ we observe domed deformation on which the phase transitions from aster at the boundary to vortex at the center, Fig.~\ref{fig:f3}d. Close to the boundary, the surface sharply deforms toward a cylindrical shape, abruptly reducing the splay at the cost of bend. On the cylindrical surface, the director field transitions from an aster to a vortex, further reducing splay and introducing additional twist and bend distortions. The surface flattens toward the top featuring a vortex. Overall, this buckling mode is able to achieve the highest degree of deformation. 

    Twist distortions are only possible in spiral regions of the defect when $\psi$ has an intermediate value, this is only observed in the deformation modes that feature a transition in the phase. Therefore the value of $k_2$ can be used to mediate the deformation of these surfaces. The splay dominated deformation 
    mode, Fig.~\ref{fig:f3}d, features twist on an extended cylindrical section of the surface. When $k_2$ is increased the cylindrical area is reduced and the spiral is pushed into the positive Gaussian curvature, Fig.~\ref{fig:f4}a\&c. This reduces the induced twist and the height of the deformation reduces asymptotically toward a finite value, cyan curve Fig.~\ref{fig:f4}a. The radial transition in the phase of the bend dominated deformation, Fig.~\ref{fig:f3}e, is entirely within the negative Gaussian curvature region, which introduces increased twist. When $k_2$ is increased, the only way to reduce the twist is to reduce the negative Gaussian curvature, which in turn reduces the magnitude of the deformation and eventually suppress this mode, magenta curve Fig.~\ref{fig:f4}a. Conversely, the conical aster deformation, Fig.~\ref{fig:f3}c, features no transition in the phase, hence no twist, and is approximately unaffected by increasing $k_2$, black curve Fig.~\ref{fig:f4}a. In fact, when $k_2$ is reduced to near zero, the director will locally distort to reduce splay and introduce twist, Fig.~\ref{fig:f4}b.
	
    We additionally observe bistability for these solutions. All surfaces shown thus far are identical under the reflection $\zeta\rightarrow-\zeta$. This symmetry can be broken by setting the boundary conditions $\zeta'(R)\neq 0$. All three previously identified buckling modes have a stable configuration in which $\zeta'$ changes sign closer to the center suggesting a similar mechanism could play a role in driving invagination, see Figs.~\ref{fig:f4}d-f. 

    Finally, this study reveals a novel mechanism for the spontaneous fusion of half-integer topological defects on fluid membranes. Consider the special case $\sigma=0$ and $k_1\!=\!k_2\!=\!k_3\!=\!k/3$. It is known that the energy minimizing nematic field with a total charge of $+1$ on a flat disc has two $+1/2$ defects. In this case, the energy scales, up to numerical pre-factors of order $1$, as ${\cal F}_f\sim \pi k/3 \log(1/\Delta)$ \cite{duclos2017topological}. However, the free-energy of a conical surface with an aster at the apex scales sub-linearly with $k$, (i.e ${\cal F}_c\sim \pi k_B \sqrt{(4k/3k_B-1)}) \log(1/\Delta)$). Therefore, a critical threshold $(k/3k_B)_c={\cal O}(1)$ arises from the balance between the energies of these two states. Thus, if $(k/3k_B)>(k/3k_B)_c$, the pair of $+1/2$ topological defects can spontaneously fuse by deforming the surface out of plane. Indeed, past works have shown that this process can occur for prescribed conical surfaces \cite{vafa2022defect}.

This mechanism for spontaneous deformation of fluid membranes can drive out-of-plane deformations of biological system, such as a film of cytoskeletal filaments on a supported lipid bilayer or a cell monolayer on a supported elastic substrate, see Sec.~6 in \cite{supp_mater}. For instance, actin and microtubule films can exhibit nematic phases \cite{sanchez2012spontaneous,sciortino2021pattern,memarian2021active}. The measured reduced Frank constant is in the range of $k=10^{-8}-10^{-7}~\mu N \mu m$ for actin filament films \cite{zhang2018interplay,yadav2019filament} and is $k=10^{-3}~\mu N \mu m$ for microtubule films \cite{velez2023probing}. Using typical values for the bending rigidity and surface tension of lipid bilayers and ignoring other possible elastic contributions, allows one to evaluate the proximity to the instability threshold~\eqref{eq:5}. In particular, in the bending dominate regime $k_B\gg \sigma R^2$, the critical threshold is of order of $(k/k_B)_c\sim 1$, and the ratio is $k/k_B=1$ for a film of actin filaments and is $k/k_B=10^4-10^5$ for a film of microtubules. In the tension dominated regime $k_B\ll \sigma R^2$, the critical threshold is $(k/\sigma R^2)_c\sim 1$ and therefore it can vary with the disc radius $R$. One can determine an upper bound by replacing the radius $R$ with the layer thickness $h$, which is the smallest lengthscale. This bound is $k/\sigma h^2<10$ for a film of both types of cytoskeletal filaments. These estimations suggest that in low-tension regimes, biological system can induce out-of-plane deformation by relieving elastic stresses at integer topological defects. In fact, this phenomenon may have been observed in Ref.~\cite{keber2014topology} that studied a film of microtubules encapsulated on a vesicle. In this work, the authors reported that lowering the surface tension leads to spindle-like vesicles with two $+1$ defects localized at the spindle poles. 

Shape dynamics driven by out-of-equilibrium processes could also be influenced by this mechanism. For instance, because the aster tends to facilitate surface deformations, it can be a nucleation point for out-of-plane deformations on active fluid membranes, like bacterial biofilms or cell monolayers  \cite{keber2014topology,maroudas2021topological,prasad2023alcanivorax,ravichandran2024topology,pearce2020defect}.

    \begin{acknowledgments}
    We are grateful to Isabelle Bonnet, Mathieu Dedenon, Karsten Kruse, Jean Fran\c cois Joanny, Jacques Prost, and Feng-Ching Tsai for insightful discussions. DJGP acknowledges funding from the Swiss National Science Foundation under starting grant TMSGI2 211367. This project has received funding from the European Union’s Horizon 2020 research and innovation programme under the Marie Skłodowska-Curie grant agreement No 847718. Q.C acknowledges funding from Institut Curie EuReCa PhD Programme. This publication reflects only the author's view and that the European Research Agency is not responsible for any use that may be made of the information it contains. 
    \end{acknowledgments}

    References cited in SI:
    \cite{keber2014topology,endresen2021topological,maroudas2021topological,blanch2021quantifying,ravichandran2024topology,bowick2009two,pearce2020defect,sanchez2012spontaneous,sciortino2021pattern,memarian2021active,evans1987physical,rawicz2000effect,roux2010membrane,phillips2009emerging,zhang2018interplay,yadav2019filament,hanakam1996myristoylated,clark2013monitoring,laplaud2021pinching,velez2023probing,oriola2020active,harris2012characterizing,da2022fibroblasts,latorre2018active,trushko2020buckling,fouchard2020curling,duque2023fracture,duclos2014perfect,saw2017topological,armengol2023epithelia,lee2011crawling,perez2019active,duclos2017topological,duclos2018spontaneous,guillamat2016probing}
    
	\bibliography{apssamp}

\begin{thebibliography}{76}%
\makeatletter
\providecommand \@ifxundefined [1]{%
 \@ifx{#1\undefined}
}%
\providecommand \@ifnum [1]{%
 \ifnum #1\expandafter \@firstoftwo
 \else \expandafter \@secondoftwo
 \fi
}%
\providecommand \@ifx [1]{%
 \ifx #1\expandafter \@firstoftwo
 \else \expandafter \@secondoftwo
 \fi
}%
\providecommand \natexlab [1]{#1}%
\providecommand \enquote  [1]{``#1''}%
\providecommand \bibnamefont  [1]{#1}%
\providecommand \bibfnamefont [1]{#1}%
\providecommand \citenamefont [1]{#1}%
\providecommand \href@noop [0]{\@secondoftwo}%
\providecommand \href [0]{\begingroup \@sanitize@url \@href}%
\providecommand \@href[1]{\@@startlink{#1}\@@href}%
\providecommand \@@href[1]{\endgroup#1\@@endlink}%
\providecommand \@sanitize@url [0]{\catcode `\\12\catcode `\$12\catcode `\&12\catcode `\#12\catcode `\^12\catcode `\_12\catcode `\%12\relax}%
\providecommand \@@startlink[1]{}%
\providecommand \@@endlink[0]{}%
\providecommand \url  [0]{\begingroup\@sanitize@url \@url }%
\providecommand \@url [1]{\endgroup\@href {#1}{\urlprefix }}%
\providecommand \urlprefix  [0]{URL }%
\providecommand \Eprint [0]{\href }%
\providecommand \doibase [0]{https://doi.org/}%
\providecommand \selectlanguage [0]{\@gobble}%
\providecommand \bibinfo  [0]{\@secondoftwo}%
\providecommand \bibfield  [0]{\@secondoftwo}%
\providecommand \translation [1]{[#1]}%
\providecommand \BibitemOpen [0]{}%
\providecommand \bibitemStop [0]{}%
\providecommand \bibitemNoStop [0]{.\EOS\space}%
\providecommand \EOS [0]{\spacefactor3000\relax}%
\providecommand \BibitemShut  [1]{\csname bibitem#1\endcsname}%
\let\auto@bib@innerbib\@empty
\bibitem [{\citenamefont {Buguin}\ \emph {et~al.}(2006)\citenamefont {Buguin}, \citenamefont {Li}, \citenamefont {Silberzan}, \citenamefont {Ladoux},\ and\ \citenamefont {Keller}}]{buguin2006micro}%
  \BibitemOpen
  \bibfield  {author} {\bibinfo {author} {\bibfnamefont {A.}~\bibnamefont {Buguin}}, \bibinfo {author} {\bibfnamefont {M.-H.}\ \bibnamefont {Li}}, \bibinfo {author} {\bibfnamefont {P.}~\bibnamefont {Silberzan}}, \bibinfo {author} {\bibfnamefont {B.}~\bibnamefont {Ladoux}},\ and\ \bibinfo {author} {\bibfnamefont {P.}~\bibnamefont {Keller}},\ }\bibfield  {title} {\bibinfo {title} {Micro-actuators: When artificial muscles made of nematic liquid crystal elastomers meet soft lithography},\ }\href@noop {} {\bibfield  {journal} {\bibinfo  {journal} {Journal of the American Chemical Society}\ }\textbf {\bibinfo {volume} {128}},\ \bibinfo {pages} {1088} (\bibinfo {year} {2006})}\BibitemShut {NoStop}%
\bibitem [{\citenamefont {Ware}\ \emph {et~al.}(2015)\citenamefont {Ware}, \citenamefont {McConney}, \citenamefont {Wie}, \citenamefont {Tondiglia},\ and\ \citenamefont {White}}]{ware2015voxelated}%
  \BibitemOpen
  \bibfield  {author} {\bibinfo {author} {\bibfnamefont {T.~H.}\ \bibnamefont {Ware}}, \bibinfo {author} {\bibfnamefont {M.~E.}\ \bibnamefont {McConney}}, \bibinfo {author} {\bibfnamefont {J.~J.}\ \bibnamefont {Wie}}, \bibinfo {author} {\bibfnamefont {V.~P.}\ \bibnamefont {Tondiglia}},\ and\ \bibinfo {author} {\bibfnamefont {T.~J.}\ \bibnamefont {White}},\ }\bibfield  {title} {\bibinfo {title} {Voxelated liquid crystal elastomers},\ }\href@noop {} {\bibfield  {journal} {\bibinfo  {journal} {Science}\ }\textbf {\bibinfo {volume} {347}},\ \bibinfo {pages} {982} (\bibinfo {year} {2015})}\BibitemShut {NoStop}%
\bibitem [{\citenamefont {Aharoni}\ \emph {et~al.}(2018)\citenamefont {Aharoni}, \citenamefont {Xia}, \citenamefont {Zhang}, \citenamefont {Kamien},\ and\ \citenamefont {Yang}}]{aharoni2018universal}%
  \BibitemOpen
  \bibfield  {author} {\bibinfo {author} {\bibfnamefont {H.}~\bibnamefont {Aharoni}}, \bibinfo {author} {\bibfnamefont {Y.}~\bibnamefont {Xia}}, \bibinfo {author} {\bibfnamefont {X.}~\bibnamefont {Zhang}}, \bibinfo {author} {\bibfnamefont {R.~D.}\ \bibnamefont {Kamien}},\ and\ \bibinfo {author} {\bibfnamefont {S.}~\bibnamefont {Yang}},\ }\bibfield  {title} {\bibinfo {title} {Universal inverse design of surfaces with thin nematic elastomer sheets},\ }\href@noop {} {\bibfield  {journal} {\bibinfo  {journal} {Proceedings of the National Academy of Sciences}\ }\textbf {\bibinfo {volume} {115}},\ \bibinfo {pages} {7206} (\bibinfo {year} {2018})}\BibitemShut {NoStop}%
\bibitem [{\citenamefont {Pikul}\ \emph {et~al.}(2017)\citenamefont {Pikul}, \citenamefont {Li}, \citenamefont {Bai}, \citenamefont {Hanlon}, \citenamefont {Cohen},\ and\ \citenamefont {Shepherd}}]{pikul2017stretchable}%
  \BibitemOpen
  \bibfield  {author} {\bibinfo {author} {\bibfnamefont {J.}~\bibnamefont {Pikul}}, \bibinfo {author} {\bibfnamefont {S.}~\bibnamefont {Li}}, \bibinfo {author} {\bibfnamefont {H.}~\bibnamefont {Bai}}, \bibinfo {author} {\bibfnamefont {R.}~\bibnamefont {Hanlon}}, \bibinfo {author} {\bibfnamefont {I.}~\bibnamefont {Cohen}},\ and\ \bibinfo {author} {\bibfnamefont {R.~F.}\ \bibnamefont {Shepherd}},\ }\bibfield  {title} {\bibinfo {title} {Stretchable surfaces with programmable 3d texture morphing for synthetic camouflaging skins},\ }\href@noop {} {\bibfield  {journal} {\bibinfo  {journal} {Science}\ }\textbf {\bibinfo {volume} {358}},\ \bibinfo {pages} {210} (\bibinfo {year} {2017})}\BibitemShut {NoStop}%
\bibitem [{\citenamefont {Si{\'e}fert}\ \emph {et~al.}(2019)\citenamefont {Si{\'e}fert}, \citenamefont {Reyssat}, \citenamefont {Bico},\ and\ \citenamefont {Roman}}]{siefert2019bio}%
  \BibitemOpen
  \bibfield  {author} {\bibinfo {author} {\bibfnamefont {E.}~\bibnamefont {Si{\'e}fert}}, \bibinfo {author} {\bibfnamefont {E.}~\bibnamefont {Reyssat}}, \bibinfo {author} {\bibfnamefont {J.}~\bibnamefont {Bico}},\ and\ \bibinfo {author} {\bibfnamefont {B.}~\bibnamefont {Roman}},\ }\bibfield  {title} {\bibinfo {title} {Bio-inspired pneumatic shape-morphing elastomers},\ }\href@noop {} {\bibfield  {journal} {\bibinfo  {journal} {Nature materials}\ }\textbf {\bibinfo {volume} {18}},\ \bibinfo {pages} {24} (\bibinfo {year} {2019})}\BibitemShut {NoStop}%
\bibitem [{\citenamefont {Dervaux}\ and\ \citenamefont {Amar}(2008)}]{dervaux2008morphogenesis}%
  \BibitemOpen
  \bibfield  {author} {\bibinfo {author} {\bibfnamefont {J.}~\bibnamefont {Dervaux}}\ and\ \bibinfo {author} {\bibfnamefont {M.~B.}\ \bibnamefont {Amar}},\ }\bibfield  {title} {\bibinfo {title} {Morphogenesis of growing soft tissues},\ }\href@noop {} {\bibfield  {journal} {\bibinfo  {journal} {Physical review letters}\ }\textbf {\bibinfo {volume} {101}},\ \bibinfo {pages} {068101} (\bibinfo {year} {2008})}\BibitemShut {NoStop}%
\bibitem [{\citenamefont {M{\"u}ller}\ \emph {et~al.}(2008)\citenamefont {M{\"u}ller}, \citenamefont {Amar},\ and\ \citenamefont {Guven}}]{muller2008conical}%
  \BibitemOpen
  \bibfield  {author} {\bibinfo {author} {\bibfnamefont {M.~M.}\ \bibnamefont {M{\"u}ller}}, \bibinfo {author} {\bibfnamefont {M.~B.}\ \bibnamefont {Amar}},\ and\ \bibinfo {author} {\bibfnamefont {J.}~\bibnamefont {Guven}},\ }\bibfield  {title} {\bibinfo {title} {Conical defects in growing sheets},\ }\href@noop {} {\bibfield  {journal} {\bibinfo  {journal} {Physical review letters}\ }\textbf {\bibinfo {volume} {101}},\ \bibinfo {pages} {156104} (\bibinfo {year} {2008})}\BibitemShut {NoStop}%
\bibitem [{\citenamefont {Modes}\ \emph {et~al.}(2010)\citenamefont {Modes}, \citenamefont {Bhattacharya},\ and\ \citenamefont {Warner}}]{modes2010disclination}%
  \BibitemOpen
  \bibfield  {author} {\bibinfo {author} {\bibfnamefont {C.~D.}\ \bibnamefont {Modes}}, \bibinfo {author} {\bibfnamefont {K.}~\bibnamefont {Bhattacharya}},\ and\ \bibinfo {author} {\bibfnamefont {M.}~\bibnamefont {Warner}},\ }\bibfield  {title} {\bibinfo {title} {Disclination-mediated thermo-optical response in nematic glass sheets},\ }\href@noop {} {\bibfield  {journal} {\bibinfo  {journal} {Physical Review E}\ }\textbf {\bibinfo {volume} {81}},\ \bibinfo {pages} {060701} (\bibinfo {year} {2010})}\BibitemShut {NoStop}%
\bibitem [{\citenamefont {Modes}\ and\ \citenamefont {Warner}(2011)}]{modes2011blueprinting}%
  \BibitemOpen
  \bibfield  {author} {\bibinfo {author} {\bibfnamefont {C.~D.}\ \bibnamefont {Modes}}\ and\ \bibinfo {author} {\bibfnamefont {M.}~\bibnamefont {Warner}},\ }\bibfield  {title} {\bibinfo {title} {Blueprinting nematic glass: Systematically constructing and combining active points of curvature for emergent morphology},\ }\href@noop {} {\bibfield  {journal} {\bibinfo  {journal} {Physical Review E}\ }\textbf {\bibinfo {volume} {84}},\ \bibinfo {pages} {021711} (\bibinfo {year} {2011})}\BibitemShut {NoStop}%
\bibitem [{\citenamefont {Si{\'e}fert}\ \emph {et~al.}(2020)\citenamefont {Si{\'e}fert}, \citenamefont {Reyssat}, \citenamefont {Bico},\ and\ \citenamefont {Roman}}]{siefert2020programming}%
  \BibitemOpen
  \bibfield  {author} {\bibinfo {author} {\bibfnamefont {E.}~\bibnamefont {Si{\'e}fert}}, \bibinfo {author} {\bibfnamefont {E.}~\bibnamefont {Reyssat}}, \bibinfo {author} {\bibfnamefont {J.}~\bibnamefont {Bico}},\ and\ \bibinfo {author} {\bibfnamefont {B.}~\bibnamefont {Roman}},\ }\bibfield  {title} {\bibinfo {title} {Programming stiff inflatable shells from planar patterned fabrics},\ }\href@noop {} {\bibfield  {journal} {\bibinfo  {journal} {Soft Matter}\ }\textbf {\bibinfo {volume} {16}},\ \bibinfo {pages} {7898} (\bibinfo {year} {2020})}\BibitemShut {NoStop}%
\bibitem [{\citenamefont {Lee}\ \emph {et~al.}(2017)\citenamefont {Lee}, \citenamefont {Kim}, \citenamefont {Kim}, \citenamefont {Hong}, \citenamefont {Ryu}, \citenamefont {Kim},\ and\ \citenamefont {Kim}}]{lee2017soft}%
  \BibitemOpen
  \bibfield  {author} {\bibinfo {author} {\bibfnamefont {C.}~\bibnamefont {Lee}}, \bibinfo {author} {\bibfnamefont {M.}~\bibnamefont {Kim}}, \bibinfo {author} {\bibfnamefont {Y.~J.}\ \bibnamefont {Kim}}, \bibinfo {author} {\bibfnamefont {N.}~\bibnamefont {Hong}}, \bibinfo {author} {\bibfnamefont {S.}~\bibnamefont {Ryu}}, \bibinfo {author} {\bibfnamefont {H.~J.}\ \bibnamefont {Kim}},\ and\ \bibinfo {author} {\bibfnamefont {S.}~\bibnamefont {Kim}},\ }\bibfield  {title} {\bibinfo {title} {Soft robot review},\ }\href@noop {} {\bibfield  {journal} {\bibinfo  {journal} {International Journal of Control, Automation and Systems}\ }\textbf {\bibinfo {volume} {15}},\ \bibinfo {pages} {3} (\bibinfo {year} {2017})}\BibitemShut {NoStop}%
\bibitem [{\citenamefont {Zhong-Can}\ and\ \citenamefont {Helfrich}(1989)}]{zhong1989bending}%
  \BibitemOpen
  \bibfield  {author} {\bibinfo {author} {\bibfnamefont {O.-Y.}\ \bibnamefont {Zhong-Can}}\ and\ \bibinfo {author} {\bibfnamefont {W.}~\bibnamefont {Helfrich}},\ }\bibfield  {title} {\bibinfo {title} {Bending energy of vesicle membranes: General expressions for the first, second, and third variation of the shape energy and applications to spheres and cylinders},\ }\href@noop {} {\bibfield  {journal} {\bibinfo  {journal} {Physical Review A}\ }\textbf {\bibinfo {volume} {39}},\ \bibinfo {pages} {5280} (\bibinfo {year} {1989})}\BibitemShut {NoStop}%
\bibitem [{\citenamefont {Lipowsky}(1991)}]{lipowsky1991conformation}%
  \BibitemOpen
  \bibfield  {author} {\bibinfo {author} {\bibfnamefont {R.}~\bibnamefont {Lipowsky}},\ }\bibfield  {title} {\bibinfo {title} {The conformation of membranes},\ }\href@noop {} {\bibfield  {journal} {\bibinfo  {journal} {Nature}\ }\textbf {\bibinfo {volume} {349}},\ \bibinfo {pages} {475} (\bibinfo {year} {1991})}\BibitemShut {NoStop}%
\bibitem [{\citenamefont {J{\"u}licher}\ and\ \citenamefont {Seifert}(1994)}]{julicher1994shape}%
  \BibitemOpen
  \bibfield  {author} {\bibinfo {author} {\bibfnamefont {F.}~\bibnamefont {J{\"u}licher}}\ and\ \bibinfo {author} {\bibfnamefont {U.}~\bibnamefont {Seifert}},\ }\bibfield  {title} {\bibinfo {title} {Shape equations for axisymmetric vesicles: a clarification},\ }\href@noop {} {\bibfield  {journal} {\bibinfo  {journal} {Physical Review E}\ }\textbf {\bibinfo {volume} {49}},\ \bibinfo {pages} {4728} (\bibinfo {year} {1994})}\BibitemShut {NoStop}%
\bibitem [{\citenamefont {Guven}\ and\ \citenamefont {V{\'a}zquez-Montejo}(2018)}]{guven2018geometry}%
  \BibitemOpen
  \bibfield  {author} {\bibinfo {author} {\bibfnamefont {J.}~\bibnamefont {Guven}}\ and\ \bibinfo {author} {\bibfnamefont {P.}~\bibnamefont {V{\'a}zquez-Montejo}},\ }\bibfield  {title} {\bibinfo {title} {The geometry of fluid membranes: Variational principles, symmetries and conservation laws},\ }\href@noop {} {\bibfield  {journal} {\bibinfo  {journal} {The Role of Mechanics in the Study of Lipid Bilayers}\ ,\ \bibinfo {pages} {167}} (\bibinfo {year} {2018})}\BibitemShut {NoStop}%
\bibitem [{\citenamefont {Frank}\ and\ \citenamefont {Kardar}(2008)}]{frank2008defects}%
  \BibitemOpen
  \bibfield  {author} {\bibinfo {author} {\bibfnamefont {J.~R.}\ \bibnamefont {Frank}}\ and\ \bibinfo {author} {\bibfnamefont {M.}~\bibnamefont {Kardar}},\ }\bibfield  {title} {\bibinfo {title} {Defects in nematic membranes can buckle into pseudospheres},\ }\href@noop {} {\bibfield  {journal} {\bibinfo  {journal} {Physical Review E}\ }\textbf {\bibinfo {volume} {77}},\ \bibinfo {pages} {041705} (\bibinfo {year} {2008})}\BibitemShut {NoStop}%
\bibitem [{\citenamefont {Napoli}\ and\ \citenamefont {Vergori}(2012{\natexlab{a}})}]{napoli2012surface}%
  \BibitemOpen
  \bibfield  {author} {\bibinfo {author} {\bibfnamefont {G.}~\bibnamefont {Napoli}}\ and\ \bibinfo {author} {\bibfnamefont {L.}~\bibnamefont {Vergori}},\ }\bibfield  {title} {\bibinfo {title} {Surface free energies for nematic shells},\ }\href@noop {} {\bibfield  {journal} {\bibinfo  {journal} {Physical Review E}\ }\textbf {\bibinfo {volume} {85}},\ \bibinfo {pages} {061701} (\bibinfo {year} {2012}{\natexlab{a}})}\BibitemShut {NoStop}%
\bibitem [{\citenamefont {Napoli}\ and\ \citenamefont {Vergori}(2012{\natexlab{b}})}]{napoli2012extrinsic}%
  \BibitemOpen
  \bibfield  {author} {\bibinfo {author} {\bibfnamefont {G.}~\bibnamefont {Napoli}}\ and\ \bibinfo {author} {\bibfnamefont {L.}~\bibnamefont {Vergori}},\ }\bibfield  {title} {\bibinfo {title} {Extrinsic curvature effects on nematic shells},\ }\href@noop {} {\bibfield  {journal} {\bibinfo  {journal} {Physical review letters}\ }\textbf {\bibinfo {volume} {108}},\ \bibinfo {pages} {207803} (\bibinfo {year} {2012}{\natexlab{b}})}\BibitemShut {NoStop}%
\bibitem [{\citenamefont {Pearce}(2022)}]{pearce2022coupling}%
  \BibitemOpen
  \bibfield  {author} {\bibinfo {author} {\bibfnamefont {D.~J.~G.}\ \bibnamefont {Pearce}},\ }\bibfield  {title} {\bibinfo {title} {Coupling the topological defect phase to the extrinsic curvature in nematic shells},\ }\href@noop {} {\bibfield  {journal} {\bibinfo  {journal} {Soft Matter}\ }\textbf {\bibinfo {volume} {18}},\ \bibinfo {pages} {5082} (\bibinfo {year} {2022})}\BibitemShut {NoStop}%
\bibitem [{\citenamefont {Segatti}\ \emph {et~al.}(2014)\citenamefont {Segatti}, \citenamefont {Snarski},\ and\ \citenamefont {Veneroni}}]{segatti2014equilibrium}%
  \BibitemOpen
  \bibfield  {author} {\bibinfo {author} {\bibfnamefont {A.}~\bibnamefont {Segatti}}, \bibinfo {author} {\bibfnamefont {M.}~\bibnamefont {Snarski}},\ and\ \bibinfo {author} {\bibfnamefont {M.}~\bibnamefont {Veneroni}},\ }\bibfield  {title} {\bibinfo {title} {Equilibrium configurations of nematic liquid crystals on a torus},\ }\href@noop {} {\bibfield  {journal} {\bibinfo  {journal} {Physical Review E}\ }\textbf {\bibinfo {volume} {90}},\ \bibinfo {pages} {012501} (\bibinfo {year} {2014})}\BibitemShut {NoStop}%
\bibitem [{\citenamefont {Jesenek}\ \emph {et~al.}(2015)\citenamefont {Jesenek}, \citenamefont {Kralj}, \citenamefont {Rosso},\ and\ \citenamefont {Virga}}]{jesenek2015defect}%
  \BibitemOpen
  \bibfield  {author} {\bibinfo {author} {\bibfnamefont {D.}~\bibnamefont {Jesenek}}, \bibinfo {author} {\bibfnamefont {S.}~\bibnamefont {Kralj}}, \bibinfo {author} {\bibfnamefont {R.}~\bibnamefont {Rosso}},\ and\ \bibinfo {author} {\bibfnamefont {E.~G.}\ \bibnamefont {Virga}},\ }\bibfield  {title} {\bibinfo {title} {Defect unbinding on a toroidal nematic shell},\ }\href@noop {} {\bibfield  {journal} {\bibinfo  {journal} {Soft matter}\ }\textbf {\bibinfo {volume} {11}},\ \bibinfo {pages} {2434} (\bibinfo {year} {2015})}\BibitemShut {NoStop}%
\bibitem [{\citenamefont {Nitschke}\ \emph {et~al.}(2018)\citenamefont {Nitschke}, \citenamefont {Nestler}, \citenamefont {Praetorius}, \citenamefont {L{\"o}wen},\ and\ \citenamefont {Voigt}}]{nitschke2018nematic}%
  \BibitemOpen
  \bibfield  {author} {\bibinfo {author} {\bibfnamefont {I.}~\bibnamefont {Nitschke}}, \bibinfo {author} {\bibfnamefont {M.}~\bibnamefont {Nestler}}, \bibinfo {author} {\bibfnamefont {S.}~\bibnamefont {Praetorius}}, \bibinfo {author} {\bibfnamefont {H.}~\bibnamefont {L{\"o}wen}},\ and\ \bibinfo {author} {\bibfnamefont {A.}~\bibnamefont {Voigt}},\ }\bibfield  {title} {\bibinfo {title} {Nematic liquid crystals on curved surfaces: a thin film limit},\ }\href@noop {} {\bibfield  {journal} {\bibinfo  {journal} {Proceedings of the Royal Society A: Mathematical, Physical and Engineering Sciences}\ }\textbf {\bibinfo {volume} {474}},\ \bibinfo {pages} {20170686} (\bibinfo {year} {2018})}\BibitemShut {NoStop}%
\bibitem [{\citenamefont {Nestler}\ \emph {et~al.}(2018)\citenamefont {Nestler}, \citenamefont {Nitschke}, \citenamefont {Praetorius},\ and\ \citenamefont {Voigt}}]{nestler2018orientational}%
  \BibitemOpen
  \bibfield  {author} {\bibinfo {author} {\bibfnamefont {M.}~\bibnamefont {Nestler}}, \bibinfo {author} {\bibfnamefont {I.}~\bibnamefont {Nitschke}}, \bibinfo {author} {\bibfnamefont {S.}~\bibnamefont {Praetorius}},\ and\ \bibinfo {author} {\bibfnamefont {A.}~\bibnamefont {Voigt}},\ }\bibfield  {title} {\bibinfo {title} {Orientational order on surfaces: The coupling of topology, geometry, and dynamics},\ }\href@noop {} {\bibfield  {journal} {\bibinfo  {journal} {Journal of Nonlinear Science}\ }\textbf {\bibinfo {volume} {28}},\ \bibinfo {pages} {147} (\bibinfo {year} {2018})}\BibitemShut {NoStop}%
\bibitem [{\citenamefont {Nestler}\ \emph {et~al.}(2020)\citenamefont {Nestler}, \citenamefont {Nitschke}, \citenamefont {L{\"o}wen},\ and\ \citenamefont {Voigt}}]{nestler2020properties}%
  \BibitemOpen
  \bibfield  {author} {\bibinfo {author} {\bibfnamefont {M.}~\bibnamefont {Nestler}}, \bibinfo {author} {\bibfnamefont {I.}~\bibnamefont {Nitschke}}, \bibinfo {author} {\bibfnamefont {H.}~\bibnamefont {L{\"o}wen}},\ and\ \bibinfo {author} {\bibfnamefont {A.}~\bibnamefont {Voigt}},\ }\bibfield  {title} {\bibinfo {title} {Properties of surface landau--de gennes q-tensor models},\ }\href@noop {} {\bibfield  {journal} {\bibinfo  {journal} {Soft Matter}\ }\textbf {\bibinfo {volume} {16}},\ \bibinfo {pages} {4032} (\bibinfo {year} {2020})}\BibitemShut {NoStop}%
\bibitem [{\citenamefont {Vafa}\ \emph {et~al.}(2022)\citenamefont {Vafa}, \citenamefont {Zhang},\ and\ \citenamefont {Nelson}}]{vafa2022defect}%
  \BibitemOpen
  \bibfield  {author} {\bibinfo {author} {\bibfnamefont {F.}~\bibnamefont {Vafa}}, \bibinfo {author} {\bibfnamefont {G.~H.}\ \bibnamefont {Zhang}},\ and\ \bibinfo {author} {\bibfnamefont {D.~R.}\ \bibnamefont {Nelson}},\ }\bibfield  {title} {\bibinfo {title} {Defect absorption and emission for p-atic liquid crystals on cones},\ }\href@noop {} {\bibfield  {journal} {\bibinfo  {journal} {Physical Review E}\ }\textbf {\bibinfo {volume} {106}},\ \bibinfo {pages} {024704} (\bibinfo {year} {2022})}\BibitemShut {NoStop}%
\bibitem [{\citenamefont {Santiago}(2018)}]{santiago2018stresses}%
  \BibitemOpen
  \bibfield  {author} {\bibinfo {author} {\bibfnamefont {J.}~\bibnamefont {Santiago}},\ }\bibfield  {title} {\bibinfo {title} {Stresses in curved nematic membranes},\ }\href@noop {} {\bibfield  {journal} {\bibinfo  {journal} {Physical Review E}\ }\textbf {\bibinfo {volume} {97}},\ \bibinfo {pages} {052706} (\bibinfo {year} {2018})}\BibitemShut {NoStop}%
\bibitem [{\citenamefont {Santiago}\ \emph {et~al.}(2019)\citenamefont {Santiago}, \citenamefont {Chac{\'o}n-Acosta},\ and\ \citenamefont {Monroy}}]{santiago2019membrane}%
  \BibitemOpen
  \bibfield  {author} {\bibinfo {author} {\bibfnamefont {J.}~\bibnamefont {Santiago}}, \bibinfo {author} {\bibfnamefont {G.}~\bibnamefont {Chac{\'o}n-Acosta}},\ and\ \bibinfo {author} {\bibfnamefont {F.}~\bibnamefont {Monroy}},\ }\bibfield  {title} {\bibinfo {title} {Membrane stress and torque induced by frank's nematic textures: A geometric perspective using surface-based constraints},\ }\href@noop {} {\bibfield  {journal} {\bibinfo  {journal} {Physical Review E}\ }\textbf {\bibinfo {volume} {100}},\ \bibinfo {pages} {012704} (\bibinfo {year} {2019})}\BibitemShut {NoStop}%
\bibitem [{\citenamefont {Pearce}\ \emph {et~al.}(2019)\citenamefont {Pearce}, \citenamefont {Ellis}, \citenamefont {Fernandez-Nieves},\ and\ \citenamefont {Giomi}}]{pearce2019geometrical}%
  \BibitemOpen
  \bibfield  {author} {\bibinfo {author} {\bibfnamefont {D.~J.~G.}\ \bibnamefont {Pearce}}, \bibinfo {author} {\bibfnamefont {P.~W.}\ \bibnamefont {Ellis}}, \bibinfo {author} {\bibfnamefont {A.}~\bibnamefont {Fernandez-Nieves}},\ and\ \bibinfo {author} {\bibfnamefont {L.}~\bibnamefont {Giomi}},\ }\bibfield  {title} {\bibinfo {title} {Geometrical control of active turbulence in curved topographies},\ }\href@noop {} {\bibfield  {journal} {\bibinfo  {journal} {Physical review letters}\ }\textbf {\bibinfo {volume} {122}},\ \bibinfo {pages} {168002} (\bibinfo {year} {2019})}\BibitemShut {NoStop}%
\bibitem [{\citenamefont {Pearce}(2020)}]{pearce2020defect}%
  \BibitemOpen
  \bibfield  {author} {\bibinfo {author} {\bibfnamefont {D.~J.~G.}\ \bibnamefont {Pearce}},\ }\bibfield  {title} {\bibinfo {title} {Defect order in active nematics on a curved surface},\ }\href@noop {} {\bibfield  {journal} {\bibinfo  {journal} {New Journal of Physics}\ }\textbf {\bibinfo {volume} {22}},\ \bibinfo {pages} {063051} (\bibinfo {year} {2020})}\BibitemShut {NoStop}%
\bibitem [{\citenamefont {Khoromskaia}\ and\ \citenamefont {Salbreux}(2023)}]{khoromskaia2023active}%
  \BibitemOpen
  \bibfield  {author} {\bibinfo {author} {\bibfnamefont {D.}~\bibnamefont {Khoromskaia}}\ and\ \bibinfo {author} {\bibfnamefont {G.}~\bibnamefont {Salbreux}},\ }\bibfield  {title} {\bibinfo {title} {Active morphogenesis of patterned epithelial shells},\ }\href@noop {} {\bibfield  {journal} {\bibinfo  {journal} {eLife}\ }\textbf {\bibinfo {volume} {12}},\ \bibinfo {pages} {e75878} (\bibinfo {year} {2023})}\BibitemShut {NoStop}%
\bibitem [{\citenamefont {Nestler}\ and\ \citenamefont {Voigt}(2021)}]{nestler2021active}%
  \BibitemOpen
  \bibfield  {author} {\bibinfo {author} {\bibfnamefont {M.}~\bibnamefont {Nestler}}\ and\ \bibinfo {author} {\bibfnamefont {A.}~\bibnamefont {Voigt}},\ }\bibfield  {title} {\bibinfo {title} {Active nematodynamics on curved surfaces--the influence of geometric forces on motion patterns of topological defects},\ }\href@noop {} {\bibfield  {journal} {\bibinfo  {journal} {arXiv preprint arXiv:2107.07779}\ } (\bibinfo {year} {2021})}\BibitemShut {NoStop}%
\bibitem [{\citenamefont {Hoffmann}\ \emph {et~al.}(2022)\citenamefont {Hoffmann}, \citenamefont {Carenza}, \citenamefont {Eckert},\ and\ \citenamefont {Giomi}}]{hoffmann2022theory}%
  \BibitemOpen
  \bibfield  {author} {\bibinfo {author} {\bibfnamefont {L.~A.}\ \bibnamefont {Hoffmann}}, \bibinfo {author} {\bibfnamefont {L.~N.}\ \bibnamefont {Carenza}}, \bibinfo {author} {\bibfnamefont {J.}~\bibnamefont {Eckert}},\ and\ \bibinfo {author} {\bibfnamefont {L.}~\bibnamefont {Giomi}},\ }\bibfield  {title} {\bibinfo {title} {Theory of defect-mediated morphogenesis},\ }\href@noop {} {\bibfield  {journal} {\bibinfo  {journal} {Science Advances}\ }\textbf {\bibinfo {volume} {8}},\ \bibinfo {pages} {eabk2712} (\bibinfo {year} {2022})}\BibitemShut {NoStop}%
\bibitem [{\citenamefont {Salbreux}\ \emph {et~al.}(2022)\citenamefont {Salbreux}, \citenamefont {J{\"u}licher}, \citenamefont {Prost},\ and\ \citenamefont {Callan-Jones}}]{salbreux2022theory}%
  \BibitemOpen
  \bibfield  {author} {\bibinfo {author} {\bibfnamefont {G.}~\bibnamefont {Salbreux}}, \bibinfo {author} {\bibfnamefont {F.}~\bibnamefont {J{\"u}licher}}, \bibinfo {author} {\bibfnamefont {J.}~\bibnamefont {Prost}},\ and\ \bibinfo {author} {\bibfnamefont {A.}~\bibnamefont {Callan-Jones}},\ }\bibfield  {title} {\bibinfo {title} {Theory of nematic and polar active fluid surfaces},\ }\href@noop {} {\bibfield  {journal} {\bibinfo  {journal} {Physical Review Research}\ }\textbf {\bibinfo {volume} {4}},\ \bibinfo {pages} {033158} (\bibinfo {year} {2022})}\BibitemShut {NoStop}%
\bibitem [{\citenamefont {Bell}\ \emph {et~al.}(2022)\citenamefont {Bell}, \citenamefont {Lin}, \citenamefont {Rupprecht},\ and\ \citenamefont {Prost}}]{bell2022active}%
  \BibitemOpen
  \bibfield  {author} {\bibinfo {author} {\bibfnamefont {S.}~\bibnamefont {Bell}}, \bibinfo {author} {\bibfnamefont {S.-Z.}\ \bibnamefont {Lin}}, \bibinfo {author} {\bibfnamefont {J.-F.}\ \bibnamefont {Rupprecht}},\ and\ \bibinfo {author} {\bibfnamefont {J.}~\bibnamefont {Prost}},\ }\bibfield  {title} {\bibinfo {title} {Active nematic flows over curved surfaces},\ }\href@noop {} {\bibfield  {journal} {\bibinfo  {journal} {Physical Review Letters}\ }\textbf {\bibinfo {volume} {129}},\ \bibinfo {pages} {118001} (\bibinfo {year} {2022})}\BibitemShut {NoStop}%
\bibitem [{\citenamefont {Vafa}\ \emph {et~al.}(2023)\citenamefont {Vafa}, \citenamefont {Nelson},\ and\ \citenamefont {Doostmohammadi}}]{vafa2023periodic}%
  \BibitemOpen
  \bibfield  {author} {\bibinfo {author} {\bibfnamefont {F.}~\bibnamefont {Vafa}}, \bibinfo {author} {\bibfnamefont {D.~R.}\ \bibnamefont {Nelson}},\ and\ \bibinfo {author} {\bibfnamefont {A.}~\bibnamefont {Doostmohammadi}},\ }\bibfield  {title} {\bibinfo {title} {Periodic orbits, pair nucleation, and unbinding of active nematic defects on cones},\ }\href@noop {} {\bibfield  {journal} {\bibinfo  {journal} {arXiv preprint arXiv:2310.06022}\ } (\bibinfo {year} {2023})}\BibitemShut {NoStop}%
\bibitem [{\citenamefont {Singha}\ \emph {et~al.}(2023)\citenamefont {Singha}, \citenamefont {Polley},\ and\ \citenamefont {Barma}}]{singha2023clustering}%
  \BibitemOpen
  \bibfield  {author} {\bibinfo {author} {\bibfnamefont {T.}~\bibnamefont {Singha}}, \bibinfo {author} {\bibfnamefont {A.}~\bibnamefont {Polley}},\ and\ \bibinfo {author} {\bibfnamefont {M.}~\bibnamefont {Barma}},\ }\bibfield  {title} {\bibinfo {title} {Clustering of lipids driven by integrin},\ }\href@noop {} {\bibfield  {journal} {\bibinfo  {journal} {Soft Matter}\ }\textbf {\bibinfo {volume} {19}},\ \bibinfo {pages} {6814} (\bibinfo {year} {2023})}\BibitemShut {NoStop}%
\bibitem [{\citenamefont {Maroudas-Sacks}\ \emph {et~al.}(2021)\citenamefont {Maroudas-Sacks}, \citenamefont {Garion}, \citenamefont {Shani-Zerbib}, \citenamefont {Livshits}, \citenamefont {Braun},\ and\ \citenamefont {Keren}}]{maroudas2021topological}%
  \BibitemOpen
  \bibfield  {author} {\bibinfo {author} {\bibfnamefont {Y.}~\bibnamefont {Maroudas-Sacks}}, \bibinfo {author} {\bibfnamefont {L.}~\bibnamefont {Garion}}, \bibinfo {author} {\bibfnamefont {L.}~\bibnamefont {Shani-Zerbib}}, \bibinfo {author} {\bibfnamefont {A.}~\bibnamefont {Livshits}}, \bibinfo {author} {\bibfnamefont {E.}~\bibnamefont {Braun}},\ and\ \bibinfo {author} {\bibfnamefont {K.}~\bibnamefont {Keren}},\ }\bibfield  {title} {\bibinfo {title} {Topological defects in the nematic order of actin fibres as organization centres of hydra morphogenesis},\ }\href@noop {} {\bibfield  {journal} {\bibinfo  {journal} {Nature Physics}\ }\textbf {\bibinfo {volume} {17}},\ \bibinfo {pages} {251} (\bibinfo {year} {2021})}\BibitemShut {NoStop}%
\bibitem [{\citenamefont {Keber}\ \emph {et~al.}(2014)\citenamefont {Keber}, \citenamefont {Loiseau}, \citenamefont {Sanchez}, \citenamefont {DeCamp}, \citenamefont {Giomi}, \citenamefont {Bowick}, \citenamefont {Marchetti}, \citenamefont {Dogic},\ and\ \citenamefont {Bausch}}]{keber2014topology}%
  \BibitemOpen
  \bibfield  {author} {\bibinfo {author} {\bibfnamefont {F.~C.}\ \bibnamefont {Keber}}, \bibinfo {author} {\bibfnamefont {E.}~\bibnamefont {Loiseau}}, \bibinfo {author} {\bibfnamefont {T.}~\bibnamefont {Sanchez}}, \bibinfo {author} {\bibfnamefont {S.~J.}\ \bibnamefont {DeCamp}}, \bibinfo {author} {\bibfnamefont {L.}~\bibnamefont {Giomi}}, \bibinfo {author} {\bibfnamefont {M.~J.}\ \bibnamefont {Bowick}}, \bibinfo {author} {\bibfnamefont {M.~C.}\ \bibnamefont {Marchetti}}, \bibinfo {author} {\bibfnamefont {Z.}~\bibnamefont {Dogic}},\ and\ \bibinfo {author} {\bibfnamefont {A.~R.}\ \bibnamefont {Bausch}},\ }\bibfield  {title} {\bibinfo {title} {Topology and dynamics of active nematic vesicles},\ }\href@noop {} {\bibfield  {journal} {\bibinfo  {journal} {Science}\ }\textbf {\bibinfo {volume} {345}},\ \bibinfo {pages} {1135} (\bibinfo {year} {2014})}\BibitemShut {NoStop}%
\bibitem [{\citenamefont {Blanch-Mercader}\ \emph {et~al.}(2021)\citenamefont {Blanch-Mercader}, \citenamefont {Guillamat}, \citenamefont {Roux},\ and\ \citenamefont {Kruse}}]{blanch2021quantifying}%
  \BibitemOpen
  \bibfield  {author} {\bibinfo {author} {\bibfnamefont {C.}~\bibnamefont {Blanch-Mercader}}, \bibinfo {author} {\bibfnamefont {P.}~\bibnamefont {Guillamat}}, \bibinfo {author} {\bibfnamefont {A.}~\bibnamefont {Roux}},\ and\ \bibinfo {author} {\bibfnamefont {K.}~\bibnamefont {Kruse}},\ }\bibfield  {title} {\bibinfo {title} {Quantifying material properties of cell monolayers by analyzing integer topological defects},\ }\href@noop {} {\bibfield  {journal} {\bibinfo  {journal} {Physical Review Letters}\ }\textbf {\bibinfo {volume} {126}},\ \bibinfo {pages} {028101} (\bibinfo {year} {2021})}\BibitemShut {NoStop}%
\bibitem [{\citenamefont {Ravichandran}\ \emph {et~al.}(2024)\citenamefont {Ravichandran}, \citenamefont {Vogg}, \citenamefont {Kruse}, \citenamefont {Pearce},\ and\ \citenamefont {Roux}}]{ravichandran2024topology}%
  \BibitemOpen
  \bibfield  {author} {\bibinfo {author} {\bibfnamefont {Y.}~\bibnamefont {Ravichandran}}, \bibinfo {author} {\bibfnamefont {M.}~\bibnamefont {Vogg}}, \bibinfo {author} {\bibfnamefont {K.}~\bibnamefont {Kruse}}, \bibinfo {author} {\bibfnamefont {D.~J.}\ \bibnamefont {Pearce}},\ and\ \bibinfo {author} {\bibfnamefont {A.}~\bibnamefont {Roux}},\ }\bibfield  {title} {\bibinfo {title} {Topology changes of the regenerating hydra define actin nematic defects as mechanical organizers of morphogenesis},\ }\href@noop {} {\bibfield  {journal} {\bibinfo  {journal} {bioRxiv}\ ,\ \bibinfo {pages} {2024}} (\bibinfo {year} {2024})}\BibitemShut {NoStop}%
\bibitem [{\citenamefont {Frank}(1958)}]{frank1958liquid}%
  \BibitemOpen
  \bibfield  {author} {\bibinfo {author} {\bibfnamefont {F.~C.}\ \bibnamefont {Frank}},\ }\bibfield  {title} {\bibinfo {title} {I. liquid crystals. on the theory of liquid crystals},\ }\href@noop {} {\bibfield  {journal} {\bibinfo  {journal} {Discussions of the Faraday Society}\ }\textbf {\bibinfo {volume} {25}},\ \bibinfo {pages} {19} (\bibinfo {year} {1958})}\BibitemShut {NoStop}%
\bibitem [{\citenamefont {De~Gennes}\ and\ \citenamefont {Prost}(1993)}]{de1993physics}%
  \BibitemOpen
  \bibfield  {author} {\bibinfo {author} {\bibfnamefont {P.}~\bibnamefont {De~Gennes}}\ and\ \bibinfo {author} {\bibfnamefont {J.}~\bibnamefont {Prost}},\ }\bibfield  {title} {\bibinfo {title} {The physics of liquid crystals, 1993},\ }\href@noop {} {\bibfield  {journal} {\bibinfo  {journal} {Oxford University Press, New York, Olbrich E., Marinov O., Davidov D., Phys. Rev. E}\ }\textbf {\bibinfo {volume} {2713}},\ \bibinfo {pages} {48} (\bibinfo {year} {1993})}\BibitemShut {NoStop}%
\bibitem [{\citenamefont {Nestler}\ \emph {et~al.}(2019)\citenamefont {Nestler}, \citenamefont {Nitschke},\ and\ \citenamefont {Voigt}}]{nestler2019finite}%
  \BibitemOpen
  \bibfield  {author} {\bibinfo {author} {\bibfnamefont {M.}~\bibnamefont {Nestler}}, \bibinfo {author} {\bibfnamefont {I.}~\bibnamefont {Nitschke}},\ and\ \bibinfo {author} {\bibfnamefont {A.}~\bibnamefont {Voigt}},\ }\bibfield  {title} {\bibinfo {title} {A finite element approach for vector-and tensor-valued surface pdes},\ }\href@noop {} {\bibfield  {journal} {\bibinfo  {journal} {Journal of Computational Physics}\ }\textbf {\bibinfo {volume} {389}},\ \bibinfo {pages} {48} (\bibinfo {year} {2019})}\BibitemShut {NoStop}%
\bibitem [{sup()}]{supp_mater}%
  \BibitemOpen
  \href@noop {} {\bibinfo {title} {Supplementary material}},\ \bibinfo {howpublished} {\url{URL_will_be_inserted_by_publisher}}\BibitemShut {NoStop}%
\bibitem [{Note1()}]{Note1}%
  \BibitemOpen
  \bibinfo {note} {Note that the boundary condition $\zeta '(R)\protect \!=\protect \!0$ enforces that the total Gaussian curvature, including the tip, is zero.}\BibitemShut {Stop}%
\bibitem [{\citenamefont {Duclos}\ \emph {et~al.}(2017)\citenamefont {Duclos}, \citenamefont {Erlenk{\"a}mper}, \citenamefont {Joanny},\ and\ \citenamefont {Silberzan}}]{duclos2017topological}%
  \BibitemOpen
  \bibfield  {author} {\bibinfo {author} {\bibfnamefont {G.}~\bibnamefont {Duclos}}, \bibinfo {author} {\bibfnamefont {C.}~\bibnamefont {Erlenk{\"a}mper}}, \bibinfo {author} {\bibfnamefont {J.-F.}\ \bibnamefont {Joanny}},\ and\ \bibinfo {author} {\bibfnamefont {P.}~\bibnamefont {Silberzan}},\ }\bibfield  {title} {\bibinfo {title} {Topological defects in confined populations of spindle-shaped cells},\ }\href@noop {} {\bibfield  {journal} {\bibinfo  {journal} {Nature Physics}\ }\textbf {\bibinfo {volume} {13}},\ \bibinfo {pages} {58} (\bibinfo {year} {2017})}\BibitemShut {NoStop}%
\bibitem [{\citenamefont {Sanchez}\ \emph {et~al.}(2012)\citenamefont {Sanchez}, \citenamefont {Chen}, \citenamefont {DeCamp}, \citenamefont {Heymann},\ and\ \citenamefont {Dogic}}]{sanchez2012spontaneous}%
  \BibitemOpen
  \bibfield  {author} {\bibinfo {author} {\bibfnamefont {T.}~\bibnamefont {Sanchez}}, \bibinfo {author} {\bibfnamefont {D.~T.}\ \bibnamefont {Chen}}, \bibinfo {author} {\bibfnamefont {S.~J.}\ \bibnamefont {DeCamp}}, \bibinfo {author} {\bibfnamefont {M.}~\bibnamefont {Heymann}},\ and\ \bibinfo {author} {\bibfnamefont {Z.}~\bibnamefont {Dogic}},\ }\bibfield  {title} {\bibinfo {title} {Spontaneous motion in hierarchically assembled active matter},\ }\href@noop {} {\bibfield  {journal} {\bibinfo  {journal} {Nature}\ }\textbf {\bibinfo {volume} {491}},\ \bibinfo {pages} {431} (\bibinfo {year} {2012})}\BibitemShut {NoStop}%
\bibitem [{\citenamefont {Sciortino}\ and\ \citenamefont {Bausch}(2021)}]{sciortino2021pattern}%
  \BibitemOpen
  \bibfield  {author} {\bibinfo {author} {\bibfnamefont {A.}~\bibnamefont {Sciortino}}\ and\ \bibinfo {author} {\bibfnamefont {A.~R.}\ \bibnamefont {Bausch}},\ }\bibfield  {title} {\bibinfo {title} {Pattern formation and polarity sorting of driven actin filaments on lipid membranes},\ }\href@noop {} {\bibfield  {journal} {\bibinfo  {journal} {Proceedings of the National Academy of Sciences}\ }\textbf {\bibinfo {volume} {118}},\ \bibinfo {pages} {e2017047118} (\bibinfo {year} {2021})}\BibitemShut {NoStop}%
\bibitem [{\citenamefont {Memarian}\ \emph {et~al.}(2021)\citenamefont {Memarian}, \citenamefont {Lopes}, \citenamefont {Schwarzendahl}, \citenamefont {Athani}, \citenamefont {Sarpangala}, \citenamefont {Gopinathan}, \citenamefont {Beller}, \citenamefont {Dasbiswas},\ and\ \citenamefont {Hirst}}]{memarian2021active}%
  \BibitemOpen
  \bibfield  {author} {\bibinfo {author} {\bibfnamefont {F.~L.}\ \bibnamefont {Memarian}}, \bibinfo {author} {\bibfnamefont {J.~D.}\ \bibnamefont {Lopes}}, \bibinfo {author} {\bibfnamefont {F.~J.}\ \bibnamefont {Schwarzendahl}}, \bibinfo {author} {\bibfnamefont {M.~G.}\ \bibnamefont {Athani}}, \bibinfo {author} {\bibfnamefont {N.}~\bibnamefont {Sarpangala}}, \bibinfo {author} {\bibfnamefont {A.}~\bibnamefont {Gopinathan}}, \bibinfo {author} {\bibfnamefont {D.~A.}\ \bibnamefont {Beller}}, \bibinfo {author} {\bibfnamefont {K.}~\bibnamefont {Dasbiswas}},\ and\ \bibinfo {author} {\bibfnamefont {L.~S.}\ \bibnamefont {Hirst}},\ }\bibfield  {title} {\bibinfo {title} {Active nematic order and dynamic lane formation of microtubules driven by membrane-bound diffusing motors},\ }\href@noop {} {\bibfield  {journal} {\bibinfo  {journal} {Proceedings of the National Academy of Sciences}\ }\textbf {\bibinfo {volume} {118}},\ \bibinfo {pages} {e2117107118} (\bibinfo {year} {2021})}\BibitemShut {NoStop}%
\bibitem [{\citenamefont {Zhang}\ \emph {et~al.}(2018)\citenamefont {Zhang}, \citenamefont {Kumar}, \citenamefont {Ross}, \citenamefont {Gardel},\ and\ \citenamefont {De~Pablo}}]{zhang2018interplay}%
  \BibitemOpen
  \bibfield  {author} {\bibinfo {author} {\bibfnamefont {R.}~\bibnamefont {Zhang}}, \bibinfo {author} {\bibfnamefont {N.}~\bibnamefont {Kumar}}, \bibinfo {author} {\bibfnamefont {J.~L.}\ \bibnamefont {Ross}}, \bibinfo {author} {\bibfnamefont {M.~L.}\ \bibnamefont {Gardel}},\ and\ \bibinfo {author} {\bibfnamefont {J.~J.}\ \bibnamefont {De~Pablo}},\ }\bibfield  {title} {\bibinfo {title} {Interplay of structure, elasticity, and dynamics in actin-based nematic materials},\ }\href@noop {} {\bibfield  {journal} {\bibinfo  {journal} {Proceedings of the National Academy of Sciences}\ }\textbf {\bibinfo {volume} {115}},\ \bibinfo {pages} {E124} (\bibinfo {year} {2018})}\BibitemShut {NoStop}%
\bibitem [{\citenamefont {Yadav}\ \emph {et~al.}(2019)\citenamefont {Yadav}, \citenamefont {Banerjee}, \citenamefont {Tabatabai}, \citenamefont {Kovar}, \citenamefont {Kim}, \citenamefont {Banerjee},\ and\ \citenamefont {Murrell}}]{yadav2019filament}%
  \BibitemOpen
  \bibfield  {author} {\bibinfo {author} {\bibfnamefont {V.}~\bibnamefont {Yadav}}, \bibinfo {author} {\bibfnamefont {D.~S.}\ \bibnamefont {Banerjee}}, \bibinfo {author} {\bibfnamefont {A.~P.}\ \bibnamefont {Tabatabai}}, \bibinfo {author} {\bibfnamefont {D.~R.}\ \bibnamefont {Kovar}}, \bibinfo {author} {\bibfnamefont {T.}~\bibnamefont {Kim}}, \bibinfo {author} {\bibfnamefont {S.}~\bibnamefont {Banerjee}},\ and\ \bibinfo {author} {\bibfnamefont {M.~P.}\ \bibnamefont {Murrell}},\ }\bibfield  {title} {\bibinfo {title} {Filament nucleation tunes mechanical memory in active polymer networks},\ }\href@noop {} {\bibfield  {journal} {\bibinfo  {journal} {Advanced functional materials}\ }\textbf {\bibinfo {volume} {29}},\ \bibinfo {pages} {1905243} (\bibinfo {year} {2019})}\BibitemShut {NoStop}%
\bibitem [{\citenamefont {V{\'e}lez-Cer{\'o}n}\ \emph {et~al.}(2023)\citenamefont {V{\'e}lez-Cer{\'o}n}, \citenamefont {Guillamat}, \citenamefont {Sagu{\'e}s},\ and\ \citenamefont {Ign{\'e}s-Mullol}}]{velez2023probing}%
  \BibitemOpen
  \bibfield  {author} {\bibinfo {author} {\bibfnamefont {I.}~\bibnamefont {V{\'e}lez-Cer{\'o}n}}, \bibinfo {author} {\bibfnamefont {P.}~\bibnamefont {Guillamat}}, \bibinfo {author} {\bibfnamefont {F.}~\bibnamefont {Sagu{\'e}s}},\ and\ \bibinfo {author} {\bibfnamefont {J.}~\bibnamefont {Ign{\'e}s-Mullol}},\ }\bibfield  {title} {\bibinfo {title} {Probing active nematics with in-situ microfabricated elastic inclusions},\ }\href@noop {} {\bibfield  {journal} {\bibinfo  {journal} {arXiv preprint arXiv:2307.11587}\ } (\bibinfo {year} {2023})}\BibitemShut {NoStop}%
\bibitem [{\citenamefont {Prasad}\ \emph {et~al.}(2023)\citenamefont {Prasad}, \citenamefont {Obana}, \citenamefont {Lin}, \citenamefont {Zhao}, \citenamefont {Sakai}, \citenamefont {Blanch-Mercader}, \citenamefont {Prost}, \citenamefont {Nomura}, \citenamefont {Rupprecht}, \citenamefont {Fattaccioli} \emph {et~al.}}]{prasad2023alcanivorax}%
  \BibitemOpen
  \bibfield  {author} {\bibinfo {author} {\bibfnamefont {M.}~\bibnamefont {Prasad}}, \bibinfo {author} {\bibfnamefont {N.}~\bibnamefont {Obana}}, \bibinfo {author} {\bibfnamefont {S.-Z.}\ \bibnamefont {Lin}}, \bibinfo {author} {\bibfnamefont {S.}~\bibnamefont {Zhao}}, \bibinfo {author} {\bibfnamefont {K.}~\bibnamefont {Sakai}}, \bibinfo {author} {\bibfnamefont {C.}~\bibnamefont {Blanch-Mercader}}, \bibinfo {author} {\bibfnamefont {J.}~\bibnamefont {Prost}}, \bibinfo {author} {\bibfnamefont {N.}~\bibnamefont {Nomura}}, \bibinfo {author} {\bibfnamefont {J.-F.}\ \bibnamefont {Rupprecht}}, \bibinfo {author} {\bibfnamefont {J.}~\bibnamefont {Fattaccioli}}, \emph {et~al.},\ }\bibfield  {title} {\bibinfo {title} {Alcanivorax borkumensis biofilms enhance oil degradation by interfacial tubulation},\ }\href@noop {} {\bibfield  {journal} {\bibinfo  {journal} {Science}\ }\textbf {\bibinfo {volume} {381}},\ \bibinfo {pages} {748} (\bibinfo {year} {2023})}\BibitemShut {NoStop}%
\bibitem [{\citenamefont {Endresen}\ \emph {et~al.}(2021)\citenamefont {Endresen}, \citenamefont {Kim}, \citenamefont {Pittman}, \citenamefont {Chen},\ and\ \citenamefont {Serra}}]{endresen2021topological}%
  \BibitemOpen
  \bibfield  {author} {\bibinfo {author} {\bibfnamefont {K.~D.}\ \bibnamefont {Endresen}}, \bibinfo {author} {\bibfnamefont {M.}~\bibnamefont {Kim}}, \bibinfo {author} {\bibfnamefont {M.}~\bibnamefont {Pittman}}, \bibinfo {author} {\bibfnamefont {Y.}~\bibnamefont {Chen}},\ and\ \bibinfo {author} {\bibfnamefont {F.}~\bibnamefont {Serra}},\ }\bibfield  {title} {\bibinfo {title} {Topological defects of integer charge in cell monolayers},\ }\href@noop {} {\bibfield  {journal} {\bibinfo  {journal} {Soft Matter}\ }\textbf {\bibinfo {volume} {17}},\ \bibinfo {pages} {5878} (\bibinfo {year} {2021})}\BibitemShut {NoStop}%
\bibitem [{\citenamefont {Bowick}\ and\ \citenamefont {Giomi}(2009)}]{bowick2009two}%
  \BibitemOpen
  \bibfield  {author} {\bibinfo {author} {\bibfnamefont {M.~J.}\ \bibnamefont {Bowick}}\ and\ \bibinfo {author} {\bibfnamefont {L.}~\bibnamefont {Giomi}},\ }\bibfield  {title} {\bibinfo {title} {Two-dimensional matter: order, curvature and defects},\ }\href@noop {} {\bibfield  {journal} {\bibinfo  {journal} {Advances in Physics}\ }\textbf {\bibinfo {volume} {58}},\ \bibinfo {pages} {449} (\bibinfo {year} {2009})}\BibitemShut {NoStop}%
\bibitem [{\citenamefont {Evans}\ and\ \citenamefont {Needham}(1987)}]{evans1987physical}%
  \BibitemOpen
  \bibfield  {author} {\bibinfo {author} {\bibfnamefont {E.}~\bibnamefont {Evans}}\ and\ \bibinfo {author} {\bibfnamefont {D.}~\bibnamefont {Needham}},\ }\bibfield  {title} {\bibinfo {title} {Physical properties of surfactant bilayer membranes: thermal transitions, elasticity, rigidity, cohesion and colloidal interactions},\ }\href@noop {} {\bibfield  {journal} {\bibinfo  {journal} {Journal of Physical Chemistry}\ }\textbf {\bibinfo {volume} {91}},\ \bibinfo {pages} {4219} (\bibinfo {year} {1987})}\BibitemShut {NoStop}%
\bibitem [{\citenamefont {Rawicz}\ \emph {et~al.}(2000)\citenamefont {Rawicz}, \citenamefont {Olbrich}, \citenamefont {McIntosh}, \citenamefont {Needham},\ and\ \citenamefont {Evans}}]{rawicz2000effect}%
  \BibitemOpen
  \bibfield  {author} {\bibinfo {author} {\bibfnamefont {W.}~\bibnamefont {Rawicz}}, \bibinfo {author} {\bibfnamefont {K.~C.}\ \bibnamefont {Olbrich}}, \bibinfo {author} {\bibfnamefont {T.}~\bibnamefont {McIntosh}}, \bibinfo {author} {\bibfnamefont {D.}~\bibnamefont {Needham}},\ and\ \bibinfo {author} {\bibfnamefont {E.}~\bibnamefont {Evans}},\ }\bibfield  {title} {\bibinfo {title} {Effect of chain length and unsaturation on elasticity of lipid bilayers},\ }\href@noop {} {\bibfield  {journal} {\bibinfo  {journal} {Biophysical journal}\ }\textbf {\bibinfo {volume} {79}},\ \bibinfo {pages} {328} (\bibinfo {year} {2000})}\BibitemShut {NoStop}%
\bibitem [{\citenamefont {Roux}\ \emph {et~al.}(2010)\citenamefont {Roux}, \citenamefont {Koster}, \citenamefont {Lenz}, \citenamefont {Sorre}, \citenamefont {Manneville}, \citenamefont {Nassoy},\ and\ \citenamefont {Bassereau}}]{roux2010membrane}%
  \BibitemOpen
  \bibfield  {author} {\bibinfo {author} {\bibfnamefont {A.}~\bibnamefont {Roux}}, \bibinfo {author} {\bibfnamefont {G.}~\bibnamefont {Koster}}, \bibinfo {author} {\bibfnamefont {M.}~\bibnamefont {Lenz}}, \bibinfo {author} {\bibfnamefont {B.}~\bibnamefont {Sorre}}, \bibinfo {author} {\bibfnamefont {J.-B.}\ \bibnamefont {Manneville}}, \bibinfo {author} {\bibfnamefont {P.}~\bibnamefont {Nassoy}},\ and\ \bibinfo {author} {\bibfnamefont {P.}~\bibnamefont {Bassereau}},\ }\bibfield  {title} {\bibinfo {title} {Membrane curvature controls dynamin polymerization},\ }\href@noop {} {\bibfield  {journal} {\bibinfo  {journal} {Proceedings of the National Academy of Sciences}\ }\textbf {\bibinfo {volume} {107}},\ \bibinfo {pages} {4141} (\bibinfo {year} {2010})}\BibitemShut {NoStop}%
\bibitem [{\citenamefont {Phillips}\ \emph {et~al.}(2009)\citenamefont {Phillips}, \citenamefont {Ursell}, \citenamefont {Wiggins},\ and\ \citenamefont {Sens}}]{phillips2009emerging}%
  \BibitemOpen
  \bibfield  {author} {\bibinfo {author} {\bibfnamefont {R.}~\bibnamefont {Phillips}}, \bibinfo {author} {\bibfnamefont {T.}~\bibnamefont {Ursell}}, \bibinfo {author} {\bibfnamefont {P.}~\bibnamefont {Wiggins}},\ and\ \bibinfo {author} {\bibfnamefont {P.}~\bibnamefont {Sens}},\ }\bibfield  {title} {\bibinfo {title} {Emerging roles for lipids in shaping membrane-protein function},\ }\href@noop {} {\bibfield  {journal} {\bibinfo  {journal} {Nature}\ }\textbf {\bibinfo {volume} {459}},\ \bibinfo {pages} {379} (\bibinfo {year} {2009})}\BibitemShut {NoStop}%
\bibitem [{\citenamefont {Hanakam}\ \emph {et~al.}(1996)\citenamefont {Hanakam}, \citenamefont {Albrecht}, \citenamefont {Eckerskorn}, \citenamefont {Matzner},\ and\ \citenamefont {Gerisch}}]{hanakam1996myristoylated}%
  \BibitemOpen
  \bibfield  {author} {\bibinfo {author} {\bibfnamefont {F.}~\bibnamefont {Hanakam}}, \bibinfo {author} {\bibfnamefont {R.}~\bibnamefont {Albrecht}}, \bibinfo {author} {\bibfnamefont {C.}~\bibnamefont {Eckerskorn}}, \bibinfo {author} {\bibfnamefont {M.}~\bibnamefont {Matzner}},\ and\ \bibinfo {author} {\bibfnamefont {G.}~\bibnamefont {Gerisch}},\ }\bibfield  {title} {\bibinfo {title} {Myristoylated and non-myristoylated forms of the ph sensor protein hisactophilin ii: intracellular shuttling to plasma membrane and nucleus monitored in real time by a fusion with green fluorescent protein.},\ }\href@noop {} {\bibfield  {journal} {\bibinfo  {journal} {The EMBO journal}\ }\textbf {\bibinfo {volume} {15}},\ \bibinfo {pages} {2935} (\bibinfo {year} {1996})}\BibitemShut {NoStop}%
\bibitem [{\citenamefont {Clark}\ \emph {et~al.}(2013)\citenamefont {Clark}, \citenamefont {Dierkes},\ and\ \citenamefont {Paluch}}]{clark2013monitoring}%
  \BibitemOpen
  \bibfield  {author} {\bibinfo {author} {\bibfnamefont {A.~G.}\ \bibnamefont {Clark}}, \bibinfo {author} {\bibfnamefont {K.}~\bibnamefont {Dierkes}},\ and\ \bibinfo {author} {\bibfnamefont {E.~K.}\ \bibnamefont {Paluch}},\ }\bibfield  {title} {\bibinfo {title} {Monitoring actin cortex thickness in live cells},\ }\href@noop {} {\bibfield  {journal} {\bibinfo  {journal} {Biophysical journal}\ }\textbf {\bibinfo {volume} {105}},\ \bibinfo {pages} {570} (\bibinfo {year} {2013})}\BibitemShut {NoStop}%
\bibitem [{\citenamefont {Laplaud}\ \emph {et~al.}(2021)\citenamefont {Laplaud}, \citenamefont {Levernier}, \citenamefont {Pineau}, \citenamefont {Roman}, \citenamefont {Barbier}, \citenamefont {S{\'a}ez}, \citenamefont {Lennon-Dum{\'e}nil}, \citenamefont {Vargas}, \citenamefont {Kruse}, \citenamefont {Du~Roure} \emph {et~al.}}]{laplaud2021pinching}%
  \BibitemOpen
  \bibfield  {author} {\bibinfo {author} {\bibfnamefont {V.}~\bibnamefont {Laplaud}}, \bibinfo {author} {\bibfnamefont {N.}~\bibnamefont {Levernier}}, \bibinfo {author} {\bibfnamefont {J.}~\bibnamefont {Pineau}}, \bibinfo {author} {\bibfnamefont {M.~S.}\ \bibnamefont {Roman}}, \bibinfo {author} {\bibfnamefont {L.}~\bibnamefont {Barbier}}, \bibinfo {author} {\bibfnamefont {P.~J.}\ \bibnamefont {S{\'a}ez}}, \bibinfo {author} {\bibfnamefont {A.-M.}\ \bibnamefont {Lennon-Dum{\'e}nil}}, \bibinfo {author} {\bibfnamefont {P.}~\bibnamefont {Vargas}}, \bibinfo {author} {\bibfnamefont {K.}~\bibnamefont {Kruse}}, \bibinfo {author} {\bibfnamefont {O.}~\bibnamefont {Du~Roure}}, \emph {et~al.},\ }\bibfield  {title} {\bibinfo {title} {Pinching the cortex of live cells reveals thickness instabilities caused by myosin ii motors},\ }\href@noop {} {\bibfield  {journal} {\bibinfo  {journal} {Science Advances}\ }\textbf {\bibinfo {volume} {7}},\ \bibinfo {pages} {eabe3640} (\bibinfo {year} {2021})}\BibitemShut {NoStop}%
\bibitem [{\citenamefont {Oriola}\ \emph {et~al.}(2020)\citenamefont {Oriola}, \citenamefont {J{\"u}licher},\ and\ \citenamefont {Brugu{\'e}s}}]{oriola2020active}%
  \BibitemOpen
  \bibfield  {author} {\bibinfo {author} {\bibfnamefont {D.}~\bibnamefont {Oriola}}, \bibinfo {author} {\bibfnamefont {F.}~\bibnamefont {J{\"u}licher}},\ and\ \bibinfo {author} {\bibfnamefont {J.}~\bibnamefont {Brugu{\'e}s}},\ }\bibfield  {title} {\bibinfo {title} {Active forces shape the metaphase spindle through a mechanical instability},\ }\href@noop {} {\bibfield  {journal} {\bibinfo  {journal} {Proceedings of the National Academy of Sciences}\ }\textbf {\bibinfo {volume} {117}},\ \bibinfo {pages} {16154} (\bibinfo {year} {2020})}\BibitemShut {NoStop}%
\bibitem [{\citenamefont {Harris}\ \emph {et~al.}(2012)\citenamefont {Harris}, \citenamefont {Peter}, \citenamefont {Bellis}, \citenamefont {Baum}, \citenamefont {Kabla},\ and\ \citenamefont {Charras}}]{harris2012characterizing}%
  \BibitemOpen
  \bibfield  {author} {\bibinfo {author} {\bibfnamefont {A.~R.}\ \bibnamefont {Harris}}, \bibinfo {author} {\bibfnamefont {L.}~\bibnamefont {Peter}}, \bibinfo {author} {\bibfnamefont {J.}~\bibnamefont {Bellis}}, \bibinfo {author} {\bibfnamefont {B.}~\bibnamefont {Baum}}, \bibinfo {author} {\bibfnamefont {A.~J.}\ \bibnamefont {Kabla}},\ and\ \bibinfo {author} {\bibfnamefont {G.~T.}\ \bibnamefont {Charras}},\ }\bibfield  {title} {\bibinfo {title} {Characterizing the mechanics of cultured cell monolayers},\ }\href@noop {} {\bibfield  {journal} {\bibinfo  {journal} {Proceedings of the National Academy of Sciences}\ }\textbf {\bibinfo {volume} {109}},\ \bibinfo {pages} {16449} (\bibinfo {year} {2012})}\BibitemShut {NoStop}%
\bibitem [{\citenamefont {Da~Costa}\ \emph {et~al.}(2022)\citenamefont {Da~Costa}, \citenamefont {Subbiah}, \citenamefont {Oh}, \citenamefont {Jeong}, \citenamefont {Na}, \citenamefont {Park}, \citenamefont {Choi},\ and\ \citenamefont {Shin}}]{da2022fibroblasts}%
  \BibitemOpen
  \bibfield  {author} {\bibinfo {author} {\bibfnamefont {A.~D.~S.}\ \bibnamefont {Da~Costa}}, \bibinfo {author} {\bibfnamefont {R.}~\bibnamefont {Subbiah}}, \bibinfo {author} {\bibfnamefont {S.~J.}\ \bibnamefont {Oh}}, \bibinfo {author} {\bibfnamefont {H.}~\bibnamefont {Jeong}}, \bibinfo {author} {\bibfnamefont {J.-I.}\ \bibnamefont {Na}}, \bibinfo {author} {\bibfnamefont {K.}~\bibnamefont {Park}}, \bibinfo {author} {\bibfnamefont {I.-S.}\ \bibnamefont {Choi}},\ and\ \bibinfo {author} {\bibfnamefont {J.~H.}\ \bibnamefont {Shin}},\ }\bibfield  {title} {\bibinfo {title} {Fibroblasts close a void in free space by a purse-string mechanism},\ }\href@noop {} {\bibfield  {journal} {\bibinfo  {journal} {ACS applied materials \& interfaces}\ }\textbf {\bibinfo {volume} {14}},\ \bibinfo {pages} {40522} (\bibinfo {year} {2022})}\BibitemShut {NoStop}%
\bibitem [{\citenamefont {Latorre}\ \emph {et~al.}(2018)\citenamefont {Latorre}, \citenamefont {Kale}, \citenamefont {Casares}, \citenamefont {G{\'o}mez-Gonz{\'a}lez}, \citenamefont {Uroz}, \citenamefont {Valon}, \citenamefont {Nair}, \citenamefont {Garreta}, \citenamefont {Montserrat}, \citenamefont {Del~Campo} \emph {et~al.}}]{latorre2018active}%
  \BibitemOpen
  \bibfield  {author} {\bibinfo {author} {\bibfnamefont {E.}~\bibnamefont {Latorre}}, \bibinfo {author} {\bibfnamefont {S.}~\bibnamefont {Kale}}, \bibinfo {author} {\bibfnamefont {L.}~\bibnamefont {Casares}}, \bibinfo {author} {\bibfnamefont {M.}~\bibnamefont {G{\'o}mez-Gonz{\'a}lez}}, \bibinfo {author} {\bibfnamefont {M.}~\bibnamefont {Uroz}}, \bibinfo {author} {\bibfnamefont {L.}~\bibnamefont {Valon}}, \bibinfo {author} {\bibfnamefont {R.~V.}\ \bibnamefont {Nair}}, \bibinfo {author} {\bibfnamefont {E.}~\bibnamefont {Garreta}}, \bibinfo {author} {\bibfnamefont {N.}~\bibnamefont {Montserrat}}, \bibinfo {author} {\bibfnamefont {A.}~\bibnamefont {Del~Campo}}, \emph {et~al.},\ }\bibfield  {title} {\bibinfo {title} {Active superelasticity in three-dimensional epithelia of controlled shape},\ }\href@noop {} {\bibfield  {journal} {\bibinfo  {journal} {Nature}\ }\textbf {\bibinfo {volume} {563}},\ \bibinfo {pages} {203} (\bibinfo {year} {2018})}\BibitemShut {NoStop}%
\bibitem [{\citenamefont {Trushko}\ \emph {et~al.}(2020)\citenamefont {Trushko}, \citenamefont {Di~Meglio}, \citenamefont {Merzouki}, \citenamefont {Blanch-Mercader}, \citenamefont {Abuhattum}, \citenamefont {Guck}, \citenamefont {Alessandri}, \citenamefont {Nassoy}, \citenamefont {Kruse}, \citenamefont {Chopard} \emph {et~al.}}]{trushko2020buckling}%
  \BibitemOpen
  \bibfield  {author} {\bibinfo {author} {\bibfnamefont {A.}~\bibnamefont {Trushko}}, \bibinfo {author} {\bibfnamefont {I.}~\bibnamefont {Di~Meglio}}, \bibinfo {author} {\bibfnamefont {A.}~\bibnamefont {Merzouki}}, \bibinfo {author} {\bibfnamefont {C.}~\bibnamefont {Blanch-Mercader}}, \bibinfo {author} {\bibfnamefont {S.}~\bibnamefont {Abuhattum}}, \bibinfo {author} {\bibfnamefont {J.}~\bibnamefont {Guck}}, \bibinfo {author} {\bibfnamefont {K.}~\bibnamefont {Alessandri}}, \bibinfo {author} {\bibfnamefont {P.}~\bibnamefont {Nassoy}}, \bibinfo {author} {\bibfnamefont {K.}~\bibnamefont {Kruse}}, \bibinfo {author} {\bibfnamefont {B.}~\bibnamefont {Chopard}}, \emph {et~al.},\ }\bibfield  {title} {\bibinfo {title} {Buckling of an epithelium growing under spherical confinement},\ }\href@noop {} {\bibfield  {journal} {\bibinfo  {journal} {Developmental cell}\ }\textbf {\bibinfo {volume} {54}},\ \bibinfo {pages} {655} (\bibinfo {year} {2020})}\BibitemShut {NoStop}%
\bibitem [{\citenamefont {Fouchard}\ \emph {et~al.}(2020)\citenamefont {Fouchard}, \citenamefont {Wyatt}, \citenamefont {Proag}, \citenamefont {Lisica}, \citenamefont {Khalilgharibi}, \citenamefont {Recho}, \citenamefont {Suzanne}, \citenamefont {Kabla},\ and\ \citenamefont {Charras}}]{fouchard2020curling}%
  \BibitemOpen
  \bibfield  {author} {\bibinfo {author} {\bibfnamefont {J.}~\bibnamefont {Fouchard}}, \bibinfo {author} {\bibfnamefont {T.~P.}\ \bibnamefont {Wyatt}}, \bibinfo {author} {\bibfnamefont {A.}~\bibnamefont {Proag}}, \bibinfo {author} {\bibfnamefont {A.}~\bibnamefont {Lisica}}, \bibinfo {author} {\bibfnamefont {N.}~\bibnamefont {Khalilgharibi}}, \bibinfo {author} {\bibfnamefont {P.}~\bibnamefont {Recho}}, \bibinfo {author} {\bibfnamefont {M.}~\bibnamefont {Suzanne}}, \bibinfo {author} {\bibfnamefont {A.}~\bibnamefont {Kabla}},\ and\ \bibinfo {author} {\bibfnamefont {G.}~\bibnamefont {Charras}},\ }\bibfield  {title} {\bibinfo {title} {Curling of epithelial monolayers reveals coupling between active bending and tissue tension},\ }\href@noop {} {\bibfield  {journal} {\bibinfo  {journal} {Proceedings of the National Academy of Sciences}\ }\textbf {\bibinfo {volume} {117}},\ \bibinfo {pages} {9377} (\bibinfo {year} {2020})}\BibitemShut {NoStop}%
\bibitem [{\citenamefont {Duque}\ \emph {et~al.}(2023)\citenamefont {Duque}, \citenamefont {Bonfanti}, \citenamefont {Fouchard}, \citenamefont {Ferber}, \citenamefont {Harris}, \citenamefont {Kabla},\ and\ \citenamefont {Charras}}]{duque2023fracture}%
  \BibitemOpen
  \bibfield  {author} {\bibinfo {author} {\bibfnamefont {J.}~\bibnamefont {Duque}}, \bibinfo {author} {\bibfnamefont {A.}~\bibnamefont {Bonfanti}}, \bibinfo {author} {\bibfnamefont {J.}~\bibnamefont {Fouchard}}, \bibinfo {author} {\bibfnamefont {E.}~\bibnamefont {Ferber}}, \bibinfo {author} {\bibfnamefont {A.}~\bibnamefont {Harris}}, \bibinfo {author} {\bibfnamefont {A.}~\bibnamefont {Kabla}},\ and\ \bibinfo {author} {\bibfnamefont {G.}~\bibnamefont {Charras}},\ }\bibfield  {title} {\bibinfo {title} {Fracture in living cell monolayers},\ }\href@noop {} {\bibfield  {journal} {\bibinfo  {journal} {BioRxiv}\ ,\ \bibinfo {pages} {2023}} (\bibinfo {year} {2023})}\BibitemShut {NoStop}%
\bibitem [{\citenamefont {Duclos}\ \emph {et~al.}(2014)\citenamefont {Duclos}, \citenamefont {Garcia}, \citenamefont {Yevick},\ and\ \citenamefont {Silberzan}}]{duclos2014perfect}%
  \BibitemOpen
  \bibfield  {author} {\bibinfo {author} {\bibfnamefont {G.}~\bibnamefont {Duclos}}, \bibinfo {author} {\bibfnamefont {S.}~\bibnamefont {Garcia}}, \bibinfo {author} {\bibfnamefont {H.}~\bibnamefont {Yevick}},\ and\ \bibinfo {author} {\bibfnamefont {P.}~\bibnamefont {Silberzan}},\ }\bibfield  {title} {\bibinfo {title} {Perfect nematic order in confined monolayers of spindle-shaped cells},\ }\href@noop {} {\bibfield  {journal} {\bibinfo  {journal} {Soft matter}\ }\textbf {\bibinfo {volume} {10}},\ \bibinfo {pages} {2346} (\bibinfo {year} {2014})}\BibitemShut {NoStop}%
\bibitem [{\citenamefont {Saw}\ \emph {et~al.}(2017)\citenamefont {Saw}, \citenamefont {Doostmohammadi}, \citenamefont {Nier}, \citenamefont {Kocgozlu}, \citenamefont {Thampi}, \citenamefont {Toyama}, \citenamefont {Marcq}, \citenamefont {Lim}, \citenamefont {Yeomans},\ and\ \citenamefont {Ladoux}}]{saw2017topological}%
  \BibitemOpen
  \bibfield  {author} {\bibinfo {author} {\bibfnamefont {T.~B.}\ \bibnamefont {Saw}}, \bibinfo {author} {\bibfnamefont {A.}~\bibnamefont {Doostmohammadi}}, \bibinfo {author} {\bibfnamefont {V.}~\bibnamefont {Nier}}, \bibinfo {author} {\bibfnamefont {L.}~\bibnamefont {Kocgozlu}}, \bibinfo {author} {\bibfnamefont {S.}~\bibnamefont {Thampi}}, \bibinfo {author} {\bibfnamefont {Y.}~\bibnamefont {Toyama}}, \bibinfo {author} {\bibfnamefont {P.}~\bibnamefont {Marcq}}, \bibinfo {author} {\bibfnamefont {C.~T.}\ \bibnamefont {Lim}}, \bibinfo {author} {\bibfnamefont {J.~M.}\ \bibnamefont {Yeomans}},\ and\ \bibinfo {author} {\bibfnamefont {B.}~\bibnamefont {Ladoux}},\ }\bibfield  {title} {\bibinfo {title} {Topological defects in epithelia govern cell death and extrusion},\ }\href@noop {} {\bibfield  {journal} {\bibinfo  {journal} {Nature}\ }\textbf {\bibinfo {volume} {544}},\ \bibinfo {pages} {212} (\bibinfo {year} {2017})}\BibitemShut {NoStop}%
\bibitem [{\citenamefont {Armengol-Collado}\ \emph {et~al.}(2023)\citenamefont {Armengol-Collado}, \citenamefont {Carenza}, \citenamefont {Eckert}, \citenamefont {Krommydas},\ and\ \citenamefont {Giomi}}]{armengol2023epithelia}%
  \BibitemOpen
  \bibfield  {author} {\bibinfo {author} {\bibfnamefont {J.-M.}\ \bibnamefont {Armengol-Collado}}, \bibinfo {author} {\bibfnamefont {L.~N.}\ \bibnamefont {Carenza}}, \bibinfo {author} {\bibfnamefont {J.}~\bibnamefont {Eckert}}, \bibinfo {author} {\bibfnamefont {D.}~\bibnamefont {Krommydas}},\ and\ \bibinfo {author} {\bibfnamefont {L.}~\bibnamefont {Giomi}},\ }\bibfield  {title} {\bibinfo {title} {Epithelia are multiscale active liquid crystals},\ }\href@noop {} {\bibfield  {journal} {\bibinfo  {journal} {Nature Physics}\ }\textbf {\bibinfo {volume} {19}},\ \bibinfo {pages} {1773} (\bibinfo {year} {2023})}\BibitemShut {NoStop}%
\bibitem [{\citenamefont {Lee}\ and\ \citenamefont {Wolgemuth}(2011)}]{lee2011crawling}%
  \BibitemOpen
  \bibfield  {author} {\bibinfo {author} {\bibfnamefont {P.}~\bibnamefont {Lee}}\ and\ \bibinfo {author} {\bibfnamefont {C.~W.}\ \bibnamefont {Wolgemuth}},\ }\bibfield  {title} {\bibinfo {title} {Crawling cells can close wounds without purse strings or signaling},\ }\href@noop {} {\bibfield  {journal} {\bibinfo  {journal} {PLoS computational biology}\ }\textbf {\bibinfo {volume} {7}},\ \bibinfo {pages} {e1002007} (\bibinfo {year} {2011})}\BibitemShut {NoStop}%
\bibitem [{\citenamefont {P{\'e}rez-Gonz{\'a}lez}\ \emph {et~al.}(2019)\citenamefont {P{\'e}rez-Gonz{\'a}lez}, \citenamefont {Alert}, \citenamefont {Blanch-Mercader}, \citenamefont {G{\'o}mez-Gonz{\'a}lez}, \citenamefont {Kolodziej}, \citenamefont {Bazellieres}, \citenamefont {Casademunt},\ and\ \citenamefont {Trepat}}]{perez2019active}%
  \BibitemOpen
  \bibfield  {author} {\bibinfo {author} {\bibfnamefont {C.}~\bibnamefont {P{\'e}rez-Gonz{\'a}lez}}, \bibinfo {author} {\bibfnamefont {R.}~\bibnamefont {Alert}}, \bibinfo {author} {\bibfnamefont {C.}~\bibnamefont {Blanch-Mercader}}, \bibinfo {author} {\bibfnamefont {M.}~\bibnamefont {G{\'o}mez-Gonz{\'a}lez}}, \bibinfo {author} {\bibfnamefont {T.}~\bibnamefont {Kolodziej}}, \bibinfo {author} {\bibfnamefont {E.}~\bibnamefont {Bazellieres}}, \bibinfo {author} {\bibfnamefont {J.}~\bibnamefont {Casademunt}},\ and\ \bibinfo {author} {\bibfnamefont {X.}~\bibnamefont {Trepat}},\ }\bibfield  {title} {\bibinfo {title} {Active wetting of epithelial tissues},\ }\href@noop {} {\bibfield  {journal} {\bibinfo  {journal} {Nature physics}\ }\textbf {\bibinfo {volume} {15}},\ \bibinfo {pages} {79} (\bibinfo {year} {2019})}\BibitemShut {NoStop}%
\bibitem [{\citenamefont {Duclos}\ \emph {et~al.}(2018)\citenamefont {Duclos}, \citenamefont {Blanch-Mercader}, \citenamefont {Yashunsky}, \citenamefont {Salbreux}, \citenamefont {Joanny}, \citenamefont {Prost},\ and\ \citenamefont {Silberzan}}]{duclos2018spontaneous}%
  \BibitemOpen
  \bibfield  {author} {\bibinfo {author} {\bibfnamefont {G.}~\bibnamefont {Duclos}}, \bibinfo {author} {\bibfnamefont {C.}~\bibnamefont {Blanch-Mercader}}, \bibinfo {author} {\bibfnamefont {V.}~\bibnamefont {Yashunsky}}, \bibinfo {author} {\bibfnamefont {G.}~\bibnamefont {Salbreux}}, \bibinfo {author} {\bibfnamefont {J.-F.}\ \bibnamefont {Joanny}}, \bibinfo {author} {\bibfnamefont {J.}~\bibnamefont {Prost}},\ and\ \bibinfo {author} {\bibfnamefont {P.}~\bibnamefont {Silberzan}},\ }\bibfield  {title} {\bibinfo {title} {Spontaneous shear flow in confined cellular nematics},\ }\href@noop {} {\bibfield  {journal} {\bibinfo  {journal} {Nature Physics}\ ,\ \bibinfo {pages} {1}} (\bibinfo {year} {2018})}\BibitemShut {NoStop}%
\bibitem [{\citenamefont {Guillamat}\ \emph {et~al.}(2016)\citenamefont {Guillamat}, \citenamefont {Ign{\'e}s-Mullol}, \citenamefont {Shankar}, \citenamefont {Marchetti},\ and\ \citenamefont {Sagu{\'e}s}}]{guillamat2016probing}%
  \BibitemOpen
  \bibfield  {author} {\bibinfo {author} {\bibfnamefont {P.}~\bibnamefont {Guillamat}}, \bibinfo {author} {\bibfnamefont {J.}~\bibnamefont {Ign{\'e}s-Mullol}}, \bibinfo {author} {\bibfnamefont {S.}~\bibnamefont {Shankar}}, \bibinfo {author} {\bibfnamefont {M.~C.}\ \bibnamefont {Marchetti}},\ and\ \bibinfo {author} {\bibfnamefont {F.}~\bibnamefont {Sagu{\'e}s}},\ }\bibfield  {title} {\bibinfo {title} {Probing the shear viscosity of an active nematic film},\ }\href@noop {} {\bibfield  {journal} {\bibinfo  {journal} {Physical review E}\ }\textbf {\bibinfo {volume} {94}},\ \bibinfo {pages} {060602} (\bibinfo {year} {2016})}\BibitemShut {NoStop}%
\end{thebibliography}%

\newpage
\onecolumngrid

\renewcommand{\thefigure}{S\arabic{figure}} 
\renewcommand{\theequation}{S\arabic{equation}}
\renewcommand{\thetable}{S\arabic{table}}

	
%
%
%

\setcounter{equation}{0}
\setcounter{figure}{0}
\setcounter{table}{0}

\section{Supplemental material:}

\section{Section 1: Theoretical description}

We consider a surface with the following free-energy
\begin{align}
	\mathcal{F}&=\int_\mathcal{A}\left\{k_B{\cal H}^2+k_1(\nabla\cdot \hat{n})^2+k_2(\hat{n}\cdot(\nabla\times \hat{n}))^2+k_3(\hat{n}\times(\nabla\times \hat{n}))^2+\sigma\right\}da. \label{eq:1} 
\end{align}
The first term is the bending energy with mean curvature ${\cal H}$, the second, third and fourth terms are the Frank free-energy associated respectively with splay, twist and bend distortions of the director field $\hat{n}$ and the fifth term is the compressional energy. In the same order, the corresponding elastic coefficients are: the bending rigidity $k_B$, the reduced Frank constant $k_1$, $k_2$, and $k_3$ and the surface tension $\sigma$. In the following, we assume that the system is deep into the nematic phase and impose that $|\hat{n}|=1$.


\subsection{Section 1.1: Lagrangian equations for surfaces of revolution with an integer topological defect}\label{section1-1}

In this subsection, we derive the Lagrangian equations for a surface of revolution with an integer topological defect at the axis of rotation. 

Being $\mathbf{r}$ the position vector, the surface is parametrised as follows,
\begin{align}
	\mathbf{r}&=(r(s)\cos(\theta),r(s)\sin(\theta),\zeta(s))\label{eq:2}
\end{align}
where $\theta$ is the azimuthal angle and $s$ is the arc-length and it satisfies the constrain $|\partial_s \mathbf{r}|=1$. Because of the previous constrain, one can define $\partial_s r=\cos(\phi(s))$ and $\partial_s \zeta=-\sin(\phi(s))$, where $\phi(s)$ is minus the angle of the tangent vector of the generatrix with respect to the radial direction. In addition, because of the symmetries of surfaces of revolution, we ignore variations in the $\theta$ coordinate. Hence, from now on, we denote by primes the derivatives with respect to the arclength $s$. The variation of the position vector \eqref{eq:2} with respect to $\theta$ and $s$ defines an orthogonal base on the tangent plane given by
\begin{subequations}\label{eq:3} 
	\begin{align}
		\partial_s\mathbf{r}&=(r'(s)\cos(\theta),r'(s)\sin(\theta),\zeta'(s))\\
		\partial_\theta\mathbf{r}&=r(s)(-\sin(\theta),\cos(\theta),0)
	\end{align}
\end{subequations}
which can be normalised as $\hat{e}_s=\partial_s\mathbf{r}$ and $\hat{e}_\theta=\partial_\theta\mathbf{r}/r(s)$. A normal to the tangent plane reads
\begin{align}
	\hat{{\cal N}}=&(-\zeta'(s)\cos(\theta),-\zeta'(s)\sin(\theta),r'(s))\label{eq:4} 
\end{align}
The components of the metric tensor $g_{\alpha\beta}$ read
\begin{align}
	g_{ss}&=\partial_s\mathbf{r}\cdot \partial_s\mathbf{r}=1~~~~g_{\theta\theta}=r^2~~~~g_{s\theta}=g_{\theta s}=0\label{eq:5} 
\end{align}
and the differential of area on the surface reads $da=\sqrt{det(g_{\alpha\beta})}ds d\theta=r ds d\theta$.

The components of the curvature tensor read
\begin{align}
	{\cal C}_{ss}&=-\partial^2_{ss}\mathbf{r}\cdot \hat{{\cal N}}=\phi'~~~~{\cal C}_{\theta\theta}=-\partial^2_{\theta\theta}\mathbf{r}\cdot \hat{{\cal N}}=-\zeta'r~~~~{\cal C}_{s\theta}={\cal C}_{\theta s}=-\partial^2_{s\theta}\mathbf{r}\cdot \hat{{\cal N}}=0\label{eq:6}
\end{align}
and the mean curvature as ${\cal H}=({\cal C}_{ss}/g_{ss}+{\cal C}_{\theta\theta}/g_{\theta\theta})/2=(\phi'-\zeta'/r)/2$. 

The variation of our tangent plane basis with respect to the parameters $\theta$ and $s$ reads 
\begin{align}
	\partial_s \hat{e}_s&=-\phi' \hat{{\cal N}}~~~~\partial_\theta \hat{e}_s=r' \hat{e}_\theta~~&~~\partial_s \hat{e}_\theta=0~~~\partial_\theta \hat{e}_\theta=\zeta' \hat{{\cal N}}-r' \hat{e}_s \label{eq:7}
\end{align}
which allow one to compute the tensor of distortions of the director field $\partial_\alpha \hat{n}_\beta$. We assume that the director field is prescribed and corresponds to that of an integer defect at the axis of rotation. In addition, we assume that the director field is constrained to be parallel to the surface of revolution (i.e. $\hat{n}\cdot\hat{{\cal N}}=0$). Therefore $\hat{n}=\cos{(\psi)}\hat{e}_s+\sin{(\psi)}\hat{e}_\theta$, where $\psi$ is the angle of the director field with respect to the direction $\hat{e}_s$. For simplicity, $\psi$ is considered uniform. In this case, $\psi=0$ corresponds to an aster, $0<\psi<\pi/2$ to a spiral, and $\psi=\pi/2$ to a vortex. Under these approximations and using the relations~\eqref{eq:7}, the divergence and the curl of the director field reads
\begin{subequations}
	\begin{align}
		\nabla\cdot\hat{n}&=(\hat{e}_s\partial_s +\frac{\hat{e}_\theta}{r}\partial_\theta)\cdot(\cos{(\psi)}\hat{e}_s+\sin{(\psi)}\hat{e}_\theta)\\
		&=\cos{(\psi)}\hat{e}_s\cdot \partial_s\hat{e}_s+\sin{(\psi)}\hat{e}_s\cdot \partial_s\hat{e}_\theta+\frac{\cos{(\psi)}}{r}\hat{e}_\theta\cdot \partial_\theta\hat{e}_s+\frac{\sin{(\psi)}}{r}\hat{e}_\theta\cdot \partial_\theta\hat{e}_\theta\\
		&=\frac{\cos{(\psi)}r'}{r}\\
		\nabla\times\hat{n}&=(\hat{e}_s\partial_s +\frac{\hat{e}_\theta}{r}\partial_r)\times(\cos{(\psi)}\hat{e}_s+\sin{(\psi)}\hat{e}_\theta)\\
		&=\cos{(\psi)}\hat{e}_s\times \partial_s\hat{e}_s+\sin{(\psi)}\hat{e}_s\times \partial_s\hat{e}_\theta+\frac{\cos{(\psi)}}{r}\hat{e}_\theta\times \partial_\theta\hat{e}_s+\frac{\sin{(\psi)}}{r}\hat{e}_\theta\times \partial_\theta\hat{e}_\theta\\
		&=\phi' \cos{(\psi)}\hat{e}_\theta+\frac{\sin{(\psi)}\zeta'}{r}\hat{e}_s+\frac{\sin{(\psi)}r'}{r}\hat{{\cal N}}
	\end{align}
\end{subequations}
Therefore the splay, twist and bend distortions of the director field in the free-energy \eqref{eq:1} read,
\begin{subequations}\label{eq:8} 
	\begin{align}
		k_1(\nabla\cdot \hat{n})^2&=k_1\left(\frac{\cos{(\psi)}r'}{r}\right)^2\\
		k_2(\hat{n}\cdot(\nabla\times \hat{n}))^2&=k_2\left(\left(\phi'+\frac{\zeta'}{r}\right)\cos{(\psi)}\sin{(\psi)}\right)^2\\
		k_3(\hat{n}\times(\nabla\times \hat{n}))^2&=k_3\left(\left(\frac{\sin{(\psi)}r'}{r}\right)^2+\left(\phi' \cos{(\psi)}^2-\frac{\sin{(\psi)}^2 \zeta'}{r}\right)^2\right)
	\end{align}
\end{subequations}
In this case, the free-energy \eqref{eq:1} takes the form
\begin{align}
\mathcal{F}&=\int_\mathcal{A}\Big\{\frac{k_B}{4}\left(\phi'-\frac{\zeta'}{r}\right)^2+k_1\left(\frac{\cos{(\psi)}r'}{r}\right)^2+k_2\left(\left(\phi'+\frac{\zeta'}{r}\right)\cos{(\psi)}\sin{(\psi)}\right)^2\nonumber
\\&+k_3\left(\left(\frac{\sin{(\psi)}r'}{r}\right)^2+\left(\phi' \cos{(\psi)}^2-\frac{\sin{(\psi)}^2 \zeta'}{r}\right)^2\right)+\sigma\Big\}r d\theta ds \nonumber\\
	&+\int_\mathcal{A}\gamma(r'-\cos{(\phi)})+\eta(\zeta'+\sin{(\phi)})ds d\theta=2\pi\int_\mathcal{A}{\cal L}[\phi,\phi',\zeta',r,r']ds\label{eq:10}
\end{align} 
where $\gamma$ and $\eta$ are two Lagrange multipliers to enforce the constraints associated with the arclength parametrisation. 

From Eq.~\eqref{eq:10}, one can identify a Lagrangian ${\cal L}$. Then the Lagrangian equations for variations of $\phi$, $r$ and $\zeta$ are given by
\begin{subequations}\label{eq:11} 
	\begin{align}
		\delta\phi:~~\frac{\partial\mathcal{L}}{\partial \phi}-\frac{d}{ds}\left(\frac{\partial\mathcal{L}}{\partial \phi'}\right)&=0\\
		\delta r:~~\frac{\partial\mathcal{L}}{\partial r}-\frac{d}{ds}\left(\frac{\partial\mathcal{L}}{\partial r'}\right)&=0\\
		\delta \zeta:~~\frac{\partial\mathcal{L}}{\partial \zeta}-\frac{d}{ds}\left(\frac{\partial\mathcal{L}}{\partial \zeta'}\right)&=0
	\end{align}
\end{subequations} 
with the boundary contributions
\begin{subequations}\label{eq:12} 
	\begin{align}
		\delta\phi:~~\frac{\partial\mathcal{L}}{\partial \phi'}&=0\\
		\delta r:~~\frac{\partial\mathcal{L}}{\partial r'}&=0\\
		\delta \zeta:~~\frac{\partial\mathcal{L}}{\partial \zeta'}&=0
	\end{align}
\end{subequations}
%
Because the Lagrangian \eqref{eq:10} is independent on the arclength parameter $s$, one can identify a Hamiltonian $H$, which is equal to a constant. When the total length is allowed to vary during the energy minimisation process, then this constant is zero. We provide the explicit forms of Eqs.~\eqref{eq:11}, \eqref{eq:12} and the Hamiltonian in Appendix $1$. 
	
	\subsection{Section 1.2: Normal force balance}

 In this section, we compute the normal force balance equation for a surface of revolution with a prescribed integer topological defect as describe in Section~\ref{section1-1}. 
 
	To obtain the normal force balance equation, we combine Eqs.~\eqref{eq:A1} and \eqref{eq:A3}. First, using Eq.~\eqref{eq:A1b} and Eq.~\eqref{eq:A3}, one can obtain 
	\begin{subequations}\label{eq:20} 
		\begin{align}
			(\bar{\eta}r'-\gamma \zeta')'&=-\gamma'\zeta'+\phi'(\bar{\eta}\zeta'+\gamma r')\nonumber \\
			&=-\Big\{\frac{k_B}{4}\left(\phi'^2-\left(\frac{\zeta'}{r}\right)^2\right)+(\phi' \cos{(\psi)})^2(k_2\sin{(\psi)}^2+k_3\cos{(\psi)}^2)\nonumber \\&-\left(\frac{\zeta' \sin{(\psi)}}{r}\right)^2(k_2\cos{(\psi)}^2+k_3\sin{(\psi)}^2)-(k_1\cos{(\psi)}^2+k_3 \sin{(\psi)}^2)\left(\frac{r'}{r}\right)^2\Big\}\left(\phi'+\frac{\zeta'}{r}\right)r\nonumber \\
			&+\sigma r \left(\phi'-\frac{\zeta'}{r}\right)+2(k_1\cos{(\psi)}^2+k_3 \sin{(\psi)}^2)\left(\frac{r' \zeta'}{r}\right)'
		\end{align}
	\end{subequations}	
	where we used the condition $\bar{\eta}'=0$ given by Eq.~\eqref{eq:A1c}. This expression allows one to re-express the derivative with respect to the arclength of Eq.~\eqref{eq:A1a} as
	\begin{align}
		0&=-\Big\{\frac{k_B}{4}\left(\phi'^2-\left(\frac{\zeta'}{r}\right)^2\right)+(\phi' \cos{(\psi)})^2(k_2\sin{(\psi)}^2+k_3\cos{(\psi)}^2)\nonumber \\&-\left(\frac{\zeta' \sin{(\psi)}}{r}\right)^2(k_2\cos{(\psi)}^2+k_3\sin{(\psi)}^2)-(k_1\cos{(\psi)}^2+k_3 \sin{(\psi)}^2)\left(\frac{r'}{r}\right)^2\Big\}\left(\phi'+\frac{\zeta'}{r}\right)r\nonumber \\
			&+\sigma r \left(\phi'-\frac{\zeta'}{r}\right)-\left(\frac{r' \zeta'}{r}\right)'\left(\frac{k_B}{2}+2\sin{(\psi)}^2(k_2\cos{(\psi)}^2+k_3\sin{(\psi)}^2)-2(k_1\cos{(\psi)}^2+k_3 \sin{(\psi)}^2)\right)\nonumber \\
		&-\left(\phi' r\right)''\left(\frac{k_B}{2}+2\cos{(\psi)}^2(k_2\sin{(\psi)}^2+k_3\cos{(\psi)}^2)\right)\label{eq:21}
	\end{align}
	Up to a scaling factor $r$, Eq.~\eqref{eq:21} is the normal force balance. Indeed, the terms proportional to the bending rigidity $k_B$ and the terms proportional to the surface tension $\sigma$ are equivalent to the normal forces found by O-Y Zhong-can and W. Helfrich. Similarly, the terms proportional to the splay, twist and bend Frank elastic constant $k_1$, $k_2$, and $k_3$ are equivalent to the normal force found by J.A. Santiago etal (2019). 

	\subsection{Section 1.3: Exact solutions of the normal force balance equation}

 In this section, we compute exact solution of the normal force balance equation and analyse their stability conditions. 
 
	In the special case where $\phi=\phi_0$ is constant and $\sigma=0$, equation \eqref{eq:21} reduces to
\begin{align}
		0&=-r\left(\frac{\sin{(\phi_0)}}{r}\right)\Big\{\left(\frac{\sin{(\phi_0)}}{r}\right)^2\left(\frac{k_B}{4}+\sin{(\psi)}^2(k_2\cos{(\psi)}^2+k_3\sin{(\psi)}^2)\right)+\nonumber\\&\left(\frac{\cos{(\phi_0)}}{r}\right)^2\left(\frac{k_B}{2}+2\sin{(\psi)}^2(k_2\cos{(\psi)}^2+k_3\sin{(\psi)}^2)-(k_1\cos{(\psi)}^2+k_3 \sin{(\psi)}^2)\right)\Big\}\label{eq:31}
	\end{align}
	which admits three solutions: the trivial solution $\phi_0=0$ corresponding to a flat disc, and two non-trivial solutions corresponding to cones with an inclination angle (complementary to the opening angle) given by
	\begin{align}
		\tan{(\phi_0)}^2&=\frac{(k_1\cos{(\psi)}^2+k_3 \sin{(\psi)}^2)-\frac{k_B}{2}-2\sin{(\psi)}^2(k_2\cos{(\psi)}^2+k_3\sin{(\psi)}^2)}{\frac{k_B}{4}+\sin{(\psi)}^2(k_2\cos{(\psi)}^2+k_3\sin{(\psi)}^2)}\label{eq:32}
	\end{align}
	The existence condition for these non-trivial solutions is that
	\begin{align}
		(k_1\cos{(\psi)}^2+k_3 \sin{(\psi)}^2)>\frac{k_B}{2}+2\sin{(\psi)}^2(k_2\cos{(\psi)}^2+k_3\sin{(\psi)}^2)\label{eq:25}
	\end{align}
	For an aster $\psi=0$, steady-state cones exists when $k_1>k_B/2$, whereas for a vortex $\psi=\pi/2$, no steady-state cone are expected. The existence of steady-state conical surfaces, for intermediate spirals $0<\psi<\pi/2$, depends on the values of the elastic constants. 

 Next we analyse the stability of these steady-state solutions by computing their total free-energy. Considering that $\phi=\phi_0$ is constant and $\sigma=0$, Eq.~\eqref{eq:10} reduces to
 \begin{align}
\mathcal{F}&=2\pi\Big\{(\zeta')^2\left(\frac{k_B}{4}+\sin{(\psi)}^2(k_2\cos{(\psi)}^2+k_3\sin{(\psi)}^2 )\right)\nonumber\\
&+(r')^2\left(k_1\cos{(\psi)}^2+k_3\sin{(\psi)}^2\right)\Big\}\frac{\log(R/a)}{r'}\label{eq:33}
\end{align} 
where we used that $dr=r' ds$, $\zeta'=-\sin{(\phi_0)}$ and $r'=\cos{(\phi_0)}$, and considered as boundary conditions that $r(s_2)=R$ and $r(s_1)=a$, where $R$ is the radius of the disc and $a$ is a microscopic length. Eq.~\eqref{eq:33} corresponds to the total free-energy of a integer defect with a constant angle $\psi$ at the tip of a conical surface with an inclination angle $\phi_0$.  

Minimization of Eq.~\eqref{eq:33} with respect to $\phi_0$ leads to same three solutions found by solving the force balance equation. Minimization of Eq.~\eqref{eq:33} with respect to $\psi$ leads to three types of topological defects: an aster $\psi=0$, a vortex $\psi=\pi/2$, and a spiral with an angle that satisfies
 \begin{align}
\tan{(\phi_0)}^2&=\frac{k_1-k_3}{2k_3 \sin{(\psi)}^2+k_2( \cos{(\psi)}^2- \sin{(\psi)}^2)}\label{eq:34}
\end{align}

The minimal surfaces with an aster defect are the flat disc with total free-energy 
 \begin{align}
\mathcal{F}/2\pi&= k_1\log(R/a)\label{eq:35}
\end{align} 
and the conical surface with elevation angle and total free-energy
 \begin{align}
\tan{(\phi_0)}^2&=\frac{k_1-k_B/2}{k_B/4}>0 \label{eq:36} \\
\mathcal{F}/2\pi&= 2\sqrt{(k_1-k_B/4)k_B/4}\log(R/a)\label{eq:37}
\end{align} 
The minimal surface with a vortex defect is a flat disc with total free-energy 
 \begin{align}
\mathcal{F}/2\pi&= k_3\log(R/a)\label{eq:38}
\end{align}
Finally, the minimal surface with the spiral defect that satisfies \eqref{eq:34} is a conical surface (except for $k_1=k_3$) with the elevation angle \eqref{eq:32} and total free-energy \eqref{eq:33}.

Stability conditions can be identified by comparing the energies of an aster on a minimal conical surface, an aster on a flat disc and a vortex on a flat disc. Respectively, these conditions read
 \begin{align}
\frac{k_1}{k_B}&=\frac{1}{2}\label{eq:39} \\
\frac{k_1}{k_B}&=\left(\frac{k_3}{k_B}\right)^2+\frac{1}{4}~~\text{with}~~\frac{k_1}{k_B}>\frac{1}{2}\label{eq:310} \\
\frac{k_1}{k_B}&=\frac{k_3}{k_B}\label{eq:39}
\end{align}

\subsection{Section 1.4: Linear stability of a flat disc with an integer topological defect to shape fluctuations}

In this section, we perform a linear stability analysis of a flat disc with an integer topological defect. Next, we discuss the parametric conditions whereby this state can become unstable as a function of the type of topological defect.

As shown above, the flat configuration is a steady state solution of the normal force balance equation \eqref{eq:21}. Here, we study its stability to small amplitude shape fluctuations. Therefore we consider $\phi=\delta\phi\ll1$ and its derivatives to be small. Note that $r'=1+{\cal O}(\delta\phi^2)$, and $\zeta'=-\delta\phi+{\cal O}(\delta\phi^2)$. 

To linear order in perturbations $\delta\phi$, the Lagrangian equations \eqref{eq:A1b} and \eqref{eq:A1c} provide a solution for the Lagrange multipliers $\gamma$ and $\bar{\eta}$. We enforce that their integration constant vanish. Inserting their expression into \eqref{eq:A1a}, one obtains the equation
\begin{align}
0&=A(r)\frac{\phi}{r}-B\left(\phi' r\right)'\label{eq:41}   
\end{align}
where $A=-(k_1\cos{(\psi)}^2+k_3\sin{(\psi)}^2)+\sigma r^2+k_B/2+2\sin{(\psi)}^2(k_2\cos{(\psi)}^2+k_3\sin{(\psi)}^2)$ and $B=k_B/2+2\cos{(\psi)}^2(k_2\sin{(\psi)}^2+k_3\cos{(\psi)}^2)>0$ is a positive coefficient. 

In the special case $\sigma=0$, the general solution of Eq.~\eqref{eq:41} reads
\begin{align}
\phi&=C \cosh\left(\sqrt{A/B}\log(r)+D)\right)\label{eq:42}    
\end{align}
where $C$ and $D$ are integration constants to be determined. Note that the solution \eqref{eq:42} has a logarithmic divergence at $r\rightarrow 0$ that can not be eliminated by a specific choice of the integration constants. When $A<0$, the solution \eqref{eq:42} becomes oscillatory which is a signature of an instability. Therefore, we identify the threshold whereby a flat disc with an integer topological defect can become unstable as $A=0$, or equivalently 
\begin{align}
k_1\cos{(\psi)}^2+k_3(\sin{(\psi)}^2-2\sin{(\psi)}^4)=2k_2 \sin{(\psi)}^2 \cos{(\psi)}^2+k_B/2\label{eq:43}
\end{align}
Note that this condition matches with the existence condition of conical surfaces \eqref{eq:25} found for $\sigma=0$. Equation \eqref{eq:43} reveals two linear mechanisms for generating out-of-plane deformations. On the one hand, in the splay dominated regime $k_1\gg k_2,k_3$, the first term on the left hand side of the stability condition~\eqref{eq:43} dominates. The critical splay Frank constant at which the flat disc becomes linearly unstable increase as the phase of the defect $\psi$ increases, and diverges for a vortex with $\psi=\pi/2$. On the other hand, in the bend dominated regime $k_3\gg k_1,k_2$, the stability condition~\eqref{eq:43} can be satisfied for a range of phase $\psi$ from $0$ to $\pi/4$. Therefore, topological defects with a phase $\psi>\pi/4$ are unable to linearly destabilize a flat disc by decreasing bend distortions. Finally, in the twist dominated regime $k_2\gg k_1,k_3$, flat discs are linearly stable irrespective of the type of topological defect. Therefore, we conclude that the aster and all spirals, except for the vortex, are able to generate spontaneous out-of-plane deformations on a flat disc via two linear mechanisms that reduce the energy associated with either splay or bend distortions by deforming the surface in the out-of-plane direction. 

In the general case $\sigma\neq0$, we estimated an approximate instability threshold by comparing the total energies of a conical surface surfaces with an aster to the total energy of a flat disc with either an aster or a vortex. In this case, we find that the instability threshold is increased by the stabilization of flat geometries by surface tension. 

 \subsection{Appendix 1: Lagrangian equations and Hamiltonian}

Then the equilibrium equations for variations of $\phi$, $r$ and $\zeta$ are given by
\begin{subequations}\label{eq:A1} 
	\begin{align}
		\delta\phi:0&=\bar{\eta}r'-\gamma \zeta'-\frac{\zeta'r'}{r}\left(\frac{k_B}{2}+2\sin{(\psi)}^2(k_2\cos{(\psi)}^2+k_3\sin{(\psi)}^2) \right)\nonumber\\&-\left(\phi' r\right)'\left(\frac{k_B}{2}+2\cos{(\psi)}^2(k_2\sin{(\psi)}^2+k_3\cos{(\psi)}^2)\right)\label{eq:A1a} \\
		\delta r:0&=\phi'^2\left(\frac{k_B}{4}+\cos{(\psi)}^2(k_2\sin{(\psi)}^2+k_3\cos{(\psi)}^2)\right)+\left(\left(\frac{r'}{r}\right)^2-\frac{2r''}{r}\right)(k_1\cos{(\psi)}^2+k_3\sin{(\psi)}^2)\nonumber\\
		&-\gamma'+\sigma-\left(\frac{\zeta'}{r}\right)^2\left(\frac{k_B}{4}+\sin{(\psi)}^2(k_2\cos{(\psi)}^2+k_3\sin{(\psi)}^2)\right)\label{eq:A1b} \\
		\delta \zeta:0&=(\bar{\eta})'\label{eq:A1c} 
	\end{align}
\end{subequations} 
with the boundary conditions
\begin{subequations}\label{eq:A2} 
	\begin{align}
		\delta\phi:~~0&=\phi' r\left(\frac{k_B}{2}+2\cos{(\psi)}^2(k_2\sin{(\psi)}^2+k_3\cos{(\psi)}^2)\right)-\zeta'\left(\frac{k_B}{2}+2(k_3-k_2)\cos{(\psi)}^2\sin{(\psi)}^2\right)\\
		\delta r:~~0&=\frac{2r'}{r}\left(k_1\cos{(\psi)}^2+k_3\sin{(\psi)}^2\right)+\gamma\\
		\delta \zeta:~~0&=\bar{\eta}
	\end{align}
\end{subequations}
where we defined $\bar{\eta}=\eta-\phi'(k_B/2+2(k_3-k_2)\sin{(\psi)}^2 \cos{(\psi)}^2)+\zeta'(k_B/2+2\sin{(\psi)}^2(k_2\cos{(\psi)}^2+k_3\sin{(\psi)}^2))/r$. 
 The Hamiltonian reads
 \begin{align}
 	H&=-{\cal L}+\phi'\frac{\partial\mathcal{L}}{\partial \phi'}+r' \frac{\partial\mathcal{L}}{\partial r'}+\zeta'\frac{\partial\mathcal{L}}{\partial \zeta'}\nonumber \\
 	&=\frac{r k_B}{4}\left(\phi'^2-\left(\frac{\zeta'}{r}\right)^2\right)+r k_1 \left(\frac{r' \cos{(\psi)}}{r}\right)^2+r k_2\cos{(\psi)}^2 \sin{(\psi)}^2\left(\phi'^2-\left(\frac{\zeta'}{r}\right)^2\right)\nonumber\\
 	&+r k_3\left(\left(\frac{r' \sin{(\psi)}}{r}\right)^2+(\phi'\cos{(\psi)}^2)^2-\left(\frac{\zeta'\sin{(\psi)}^2}{r}\right)^2\right)-\sigma r+\bar{\eta}\zeta'+\gamma r'\label{eq:A3}
 \end{align}
 
\section{Section 2: Height of a conical surface with arbitrary phase}

\subsection{Section 2.1: Energy of a conical surface}

We assume a cylindrical Monge gauge with $\zeta = mr$ and $\psi$ constant.  The mean curvature of such a surface is given by ${\cal H} = \frac{m}{2r\sqrt{1+m^2}}$, the area element of the surface is given by $r\sqrt{1+m^2}$. We take the radius of the membrane to be $R=1$.

With this we can write down the free-energy associated with the shape of the surface as 
\begin{align}
    F_s &= 2\pi\int_\Delta^1[k_B {\cal H}^2 + \sigma]r\sqrt{1+m^2}\textrm{d}r\\
    &= \frac{\pi k_Bm^2}{2\sqrt{1+m^2}}\log(1/\Delta) + \sigma\pi\sqrt{1+m^2}[1-\Delta^2],
\end{align}
where we have introduced $\Delta$ as the inner radius of the surface. This is to account for the fact that on a cone $\cal H$ diverges at $r=0$. For comparison with numerical results, $\Delta = 1/N$, where $N$ is the number of discretization bins used to evaluate the functions.

The energy associated with distortions in the nematic director is given by
\begin{align}
F_{LC} &= 2\pi\int_\Delta^1 \frac{k_1\cos^2(\psi)}{r\sqrt{1+m^2}} + \frac{k_2m^2\sin^2(\psi)\cos^2(\psi)}{r\sqrt{1+m^2}} + \frac{k_3\sin^2(\psi)(1+m^2\sin^2(\psi))}{r\sqrt{1+m^2}} \textrm{d}r\\
&= \frac{2\pi\log(1/\Delta)}{\sqrt{1+m^2}}\left[k_1\cos^2(\psi) + k_2m^2\sin^2(\psi)\cos^2(\psi) + k_3\sin^2(\psi)(1+m^2\sin^2(\psi))\right]
\end{align}

We introduce $L=2\pi\log(1/\Delta)$ and $M=\pi[1-\Delta^2]$. We now write the total free-energy of a conical surface with a defect of constant phase $\psi$ as 

\begin{multline}
F = L\left[k_1\cos^2(\psi) + k_3\sin^2(\psi)\right]/\sqrt{1+m^2}\\
+ L\left[k_2\sin^2(\psi)\cos^2(\psi) + k_3\sin^4(\psi) + k_B/4\right]m^2/\sqrt{1+m^2} + M\sigma\sqrt{1+m^2}\label{eq:FSI}
\end{multline}

\subsection{Section 2.2: Equilibrium height of a conical surface}

Taking the derivative with respect to $m$ and setting it to zero, we arrive at an expression for the height of the cone given by
\begin{equation}
m^2 = \frac{L\left[k_1\cos^2(\psi) + k_3\sin^2(\psi) - 2k_2\sin^2(\psi)\cos^2(\psi) - 2k_3\sin^4(\psi) - k_B/2\right]-M\sigma}{L\left[k_2\sin^2(\psi)\cos^2(\psi) + k_3\sin^4(\psi) + k_B/4\right] + M\sigma}\label{eq:m2SI}
\end{equation}

\subsection{2.3: Transitions between deformations}

\textbf{Transition to conical surface in single elastic constant limit}

In the single constant limit, we set $k_1=k_2=k_3=k/3$ and we can re-write Eq.~\ref{eq:m2SI} as 
\begin{equation}
    m^2 = \frac{(4k/3k_B)\left[1 - 2\sin^2(\psi)\right] - 2-4M\sigma/Lk_B}{(4k/3k_B)\sin^2(\psi) + 1 + 4M\sigma/k_BL}
\end{equation}

\begin{figure}[h]
	\centering
	\includegraphics[width=0.85\columnwidth]{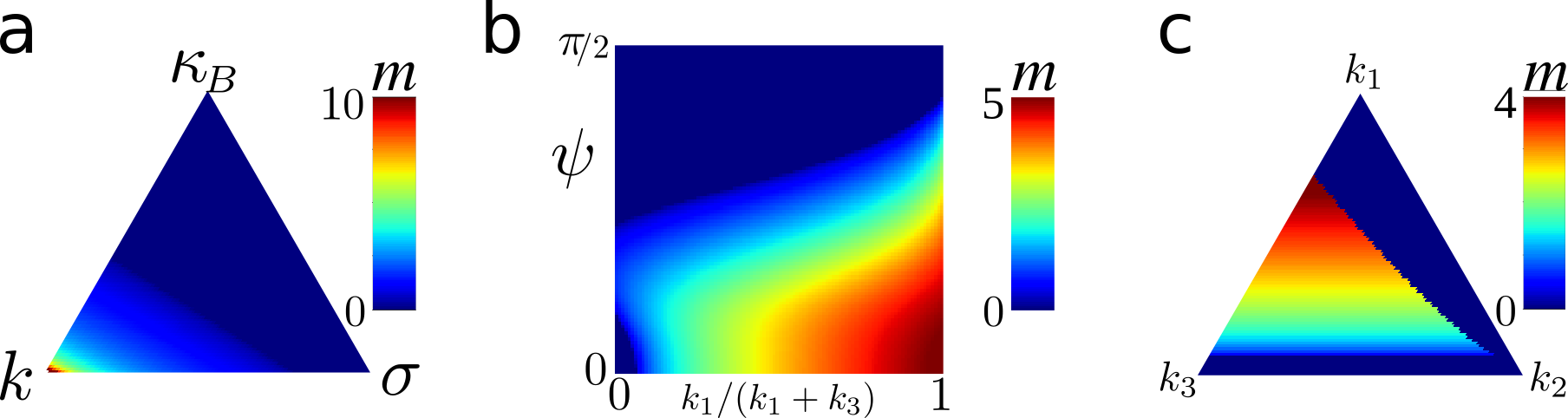}
	\caption{\label{fig:fs1} Result of Eq.~\ref{eq:m2SI} for three sets of variables. (a) Single Frank elastic constant ($k_1=k_2=k_3=k/3$) with and aster ($\psi=0$) varying $\sigma$, $k$ and $k_B$ under the constraint $k+k_B+\sigma=1$. This is analogous to Fig.1b (main text). (b) Varying Frank elastic constants and predetermined phase with $k_B=\sigma = 1/10$, $k_2=0$, $k_1+k_3=1$. This is analogous to Fig.1c (main text). (c) Varying Frank elastic constants with aster defect $\psi=0$ with $k_B=\sigma = 1/10$, $k_1+k_2+k_3=1$. This is analogous to Fig.2a (main text).}
\end{figure}

The results of this are shown in Fig.~\ref{fig:fs1}a, which is comparable to Fig 1b in the main text. 

If we set $A = 4k/3k_B$ and $B=4M\sigma/Lk_B$ we arrive at Eq.~3 in the main text. If we consider an aster, $\psi=0$, and set $m=0$ we identify the point at which the conical surface becomes stable in the single constant limit, which gives Eq.~4 and the red line on Fig.~1b in the main text.

\textbf{Transition to conical surface with variable elastic constants and predetermined phase}

For a given set of elastic coefficients and phase we calculate $m^2$ from Eq.~\ref{eq:m2SI} (if $m^2 < 0$ we set $m=0$). This is plotted in Fig.~\ref{fig:fs1}b. The green line on Fig.~1c shows where $m^2=0$ according to Eq.~\ref{eq:m2SI}.

\textbf{Transition to conical surface with variable elastic constants and variable phase}

We assume a conical aster configuration, $\psi=0$, thus Eq.~\ref{eq:m2SI} becomes
\begin{equation}
m^2 = \frac{L\left[k_1 - k_B/2\right]-M\sigma}{Lk_B/4 + M\sigma}\label{eq:SMend}
\end{equation}
The results of this equation are plotted in Fig.~\ref{fig:fs1}c. To obtain the transition curves to either an aster or a vortex on a flat disc. We then substitute the solution of Eq.~\ref{eq:SMend} into Eq.~\ref{eq:FSI} to get the energy of a conical surface with an aster. 

The conical aster energy is then compared to the energy of an aster or vortex on a flat disc, which are given by:
\begin{align}
    F_{\textrm{aster,flat}} &= L k_1+ M\sigma\\
    F_{\textrm{vortex,flat}} &= L k_3 + M\sigma
\end{align}

This gives the transition curves in Fig.~\ref{fig:fs1}c, which correspond to Eq. 4 and Eq. 5 in the main text, respectively.

\subsection{Section 2.4: The limit $\Delta\rightarrow 0$}

We describe the elastic energy of the nematic field to be given by the Frank free energy, which diverges at the core of the defect. Since we do not model the physics of the core of the defect, this necessitates a finite defect core radius, $\Delta$. Thus, we restrict our view to systems which are much bigger than the defect core radius, i.e. $\lim \Delta\rightarrow 0$. 

\begin{figure}[h]
	\centering
	\includegraphics[width=0.85\columnwidth]{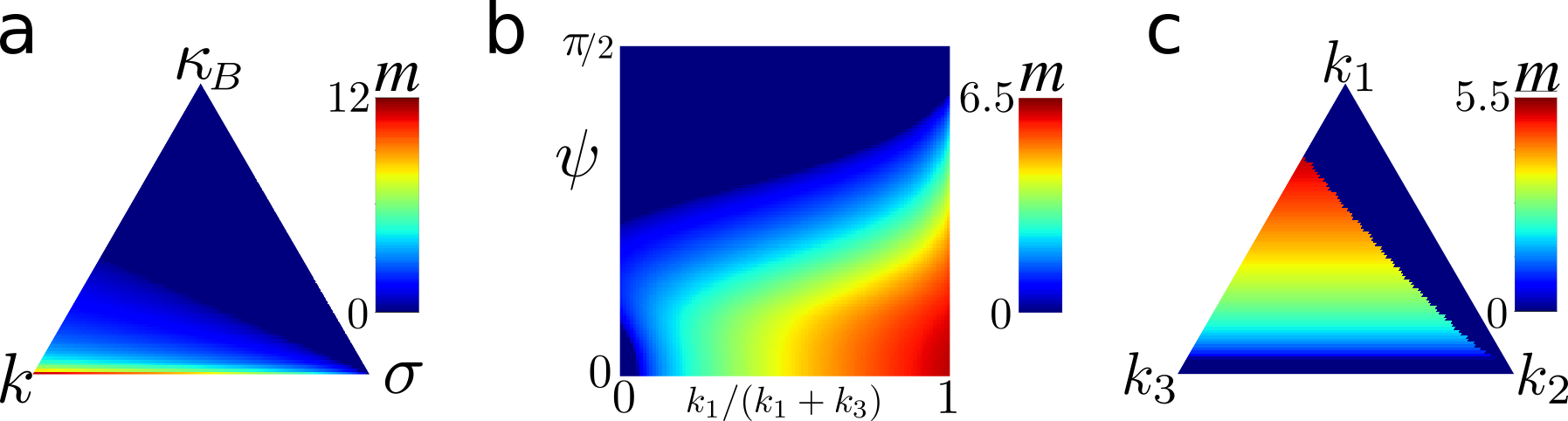}
	\caption{\label{fig:fs2} Result of Eq.~\ref{eq:m2SI} for three sets of variables evaluated for $\Delta = 0$. (a) Single Frank elastic constant ($k_1=k_2=k_3=k/3$) with and aster ($\psi=0$) varying $\sigma$, $k$ and $k_B$ under the constraint $k+k_B+\sigma=1$. This is analogous to Fig.1b (main text). (b) Varying Frank elastic constants and predetermined phase with $k_B=\sigma = 1/10$, $k_2=0$, $k_1+k_3=1$. This is analogous to Fig.1c (main text). (c) Varying Frank elastic constants with aster defect $\psi=0$ with $k_B=\sigma = 1/10$, $k_1+k_2+k_3=1$. This is analogous to Fig.2a (main text).}
\end{figure}

In the conical calculation above, Eq.~\ref{eq:FSI}, $\Delta$ modulates the relative importance of the surface tension, $\sigma$. We can see this by rescaling the energy by the defect core energy, $L=2\pi\log(1/\Delta)$. In the limit $\Delta\rightarrow 0$, the surface tension becomes insignificant relative to the other contributing energies. This allows us to evaluate all transitions in the $\Delta = 0$ limit by considering the case where $\sigma=0$. This shows very little change with respect to our previous calculations, see Fig.~\ref{fig:fs2}. There are two noticeable differences: first, the transition between buckled and flat asters in the single Frank constant limit is now independent of the value of $\sigma$. Second, the magnitude of the buckled surfaces is generally higher, this is because there is more energy available to drive the deformation. 

\subsection{Section 2.5: Effect of non-zero $\Delta$}

We use a value of $\Delta = 10^{-2}R$ in our numerical calculations, as a balance between accuracy and computational feasibility. To assess the validity of this choice, we recreate the results from Fig.~\ref{fig:fs1} for a large range of $\Delta$.

\begin{figure}[h]
	\centering
	\includegraphics[width=0.85\columnwidth]{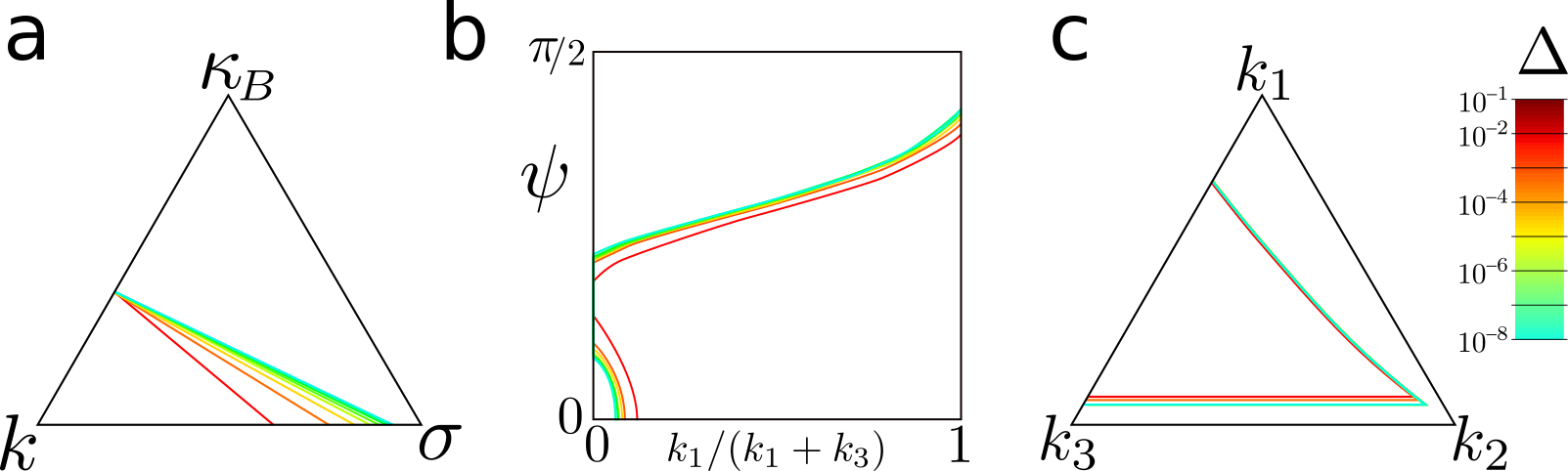}
	\caption{\label{fig:fs3} Position of the boundaries between buckled and flat sheets according to Eq.~\ref{eq:FSI} for values of $\Delta$ that vary over 7 orders of magnitude for three sets of variables. (a) Single Frank elastic constant ($k_1=k_2=k_3=k/3$) with and aster ($\psi=0$) varying $\sigma$, $k$ and $k_B$ under the constraint $k+k_B+\sigma=1$. This is analogous to Fig.1b (main text). (b) Varying Frank elastic constants and predetermined phase with $k_B=\sigma = 1/10$, $k_2=0$, $k_1+k_3=1$. This is analogous to Fig.1c (main text). (c) Varying Frank elastic constants with aster defect $\psi=0$ with $k_B=\sigma = 1/10$, $k_1+k_2+k_3=1$. This is analogous to Fig.2a (main text). }
\end{figure}

Fig.~\ref{fig:fs3} shows how the boundaries between buckled and flat sheets varies as $\Delta$ is varied over 7 orders of magnitude. It is clear here that the effect is small as $\Delta$ decreases and that our value of $\Delta = 10^{-2}R$ does not introduce significant in the transitions present. 

\section{Section 3: Minimizing the energy through Monte-Carlo method}

\subsection{Section 3.1: Evaluating the energy}

To solve using Monte-Carlo methods, the equations are recast using a cylindrical Monge gauge. The surface is given by the equations 
\begin{align}
    X&=r \textrm{cos}(\theta),\\
    Y&=r \textrm{sin}(\theta),\\
    Z&=\zeta(r)
\end{align}
and the director field is given by the function $\psi(r)$. To fully evaluate all terms of the free-energy density on the surface, we need to know the function $\psi$ and its first derivative and $\zeta$ and its first two derivatives. 

We do this by discretizing the functions in the radial direction and set the radius of the patch $R=1$ in all simulations. The space is divided into $N$ discretized points, and the locations of the center of the first and last bins are at $\epsilon = 1/2N$ and $1-\epsilon$, respectively. There are two boundaries that we need to accommodate. 

\subsubsection{Section 3.1.1: Center of the surface}

The center of the surface we assume is given by a spherical cap. At this point the energy associated with the director field diverges, and we consider the spherical cap to include no orientational order. Thus the only terms we need to calculate are those associated with the bending energy and tension of the surface. These are given by 
\begin{equation}
    F_{tip} = \int k_B {\cal H}^2 + \sigma dA
\end{equation}

The surface area of a spherical cap is given by $A = 2\pi L^2(1-\cos(\alpha))$, where $\alpha$ is the angular size of the spherical cap and $L$ is its radius. The spherical cap extends to a radius $\epsilon$ in the cylindrical radial coordinate, which is the position of the first discretization point, at which point it must smoothly transition into the function $\zeta$. After some trigonometry, we can arrive at
\begin{equation}
    F_{tip} = 2\pi k_B\left[1-\frac{1}{\sqrt{\zeta'^2(\epsilon)+1}}\right] + 2\pi\sigma \epsilon^2\left[1-\frac{1}{\sqrt{\zeta'^2(\epsilon)+1}}\right]\frac{(1+\zeta'^2(\epsilon))}{\zeta'^2(\epsilon)}
\end{equation}

\subsubsection{Section 3.1.2: Edge of the surface}

The boundary conditions at the edge of the patch are that $\zeta'(r\geq1)=M$. This is accommodated within the liquid crystal energy and surface tension by including an imaginary point beyond the boundary with the correct boundary conditions. However, for the curvature terms, which rely specifically on the second derivative of the height function, more care is required. This is to prevent a diverging value of $\zeta''$ below the lengthscale of the discretization of the surface.

We define the second derivative of the height as 
\begin{equation}
\zeta''(1) = (M-\zeta'(1-\epsilon))/\epsilon
\end{equation}
Here $\epsilon = \frac{1}{2N}$ where $N$ is the number of discretized points in $r$. Thus $\zeta'(1-\epsilon)$ is the evaluated at the last grid point, which is a distance $\epsilon$ from the boundary.

The area element at this point is given by $J = \sqrt{1+M^2}$ thus we can write the mean curvature at this point as 

\begin{equation}
{\cal H} = \frac{M+M^3 + (M-\zeta'(1-\epsilon))/\epsilon}{2J^3}
\end{equation}

We consider $\cal H$ to be constant in this small region close to the boundary which has an area equal to $2\pi\epsilon J$. Which results in an additional energy given by
\begin{equation}
F_{\rm{edge}} = 2\pi k_B\epsilon\frac{(M+M^3 + (M-\zeta'(1-\epsilon))/\epsilon)^2}{4J^5}
\end{equation}

\subsection{Section 3.2: Minimizing the energy through Monte Carlo approach}

We assume that the radial height profile of the surface can be approximated by an infinite series of Gaussian functions. This is true for any $L^1$ function (see Yves Myers'  "Wavelents and Operators" for a proof). Thus we assume that the function $\zeta$ can be written

\begin{equation}
    \zeta(r) = \sum_i A_i\exp[-(r-\mu_i)^2/2\sigma^2_i].
\end{equation}
Where $A_i$, $\sigma^2_i$ and $\mu_i$ are the amplitude, variance and mean of the constituent Gaussian functions, respectively. 

From this we introduce the perturbation 
\begin{equation}
    \Delta\zeta(r) = A\exp[-(r-\mu)^2/2\sigma^2] - A\exp[-(1-\mu)^2/2\sigma^2].
\end{equation}
We have introduced a constant here that fixes $\zeta(1) = 0$. 

In the case where the gradient at the boundary is fixed such that $\zeta'(r\geq1) = M$ we modify the perturbation to satisfy this boundary condition
\begin{equation}
    \Delta\zeta(r) = A\exp[-(r-\mu)^2/2\sigma^2] + A\exp[-(r-2+\mu)^2/2\sigma^2]- A\exp[-(1-\mu)^2/2\sigma^2]- A\exp[-(1-2+\mu)^2/2\sigma^2].
\end{equation}
This perturbation has both $\Delta\zeta(1)=0$ and $\Delta\zeta'(1)=0$, thus when combined with the initial condition $\zeta(r) = Mr$, the boundary conditions are preserved. It should be noted that this form does not strictly enforce the boundary conditions due to the discretized nature of the computational approach, it is merely used to speed up the Monte Carlo algorithm. The boundary is constrained by the additional edge energy introduced above. 

We take a similar approach with the orientation field. We assume the orientation field can be expressed
\begin{equation}
    \psi(r) = \psi_1 + \sum_i B_i\exp[-(r-\rho_i)^2/2\Sigma^2_i].
\end{equation}

We use the following perturbation $\Delta \psi$
\begin{equation}
    \Delta\psi(r) = B\exp[-(r-\rho)^2/2\Sigma^2].
\end{equation}

Similar to previous, when the boundary conditions on the director field are given by $\psi(r\geq1)=\psi_1$ we use the perturbation 
\begin{equation}
    \Delta\psi(r) = B\exp[-(r-\rho)^2/2\Sigma^2] + B\exp[-(r-2+\rho)^2/2\Sigma^2]- B\exp[-(1-\rho)^2/2\Sigma^2]- B\exp[-(1-2+\rho)^2/2\Sigma^2].
\end{equation}
This additionally fixes $\Delta\psi(1) =\Delta\psi'(1) = 0$. As before, this is not strictly necessary but is used to increase the speed of the Monte Carlo algorithm. 

These perturbations are applied sequentially to the functions $\zeta$ and $\psi$ and are accepted or rejected according to a Boltzmann factor with a gradually reducing temperature in a standard simulated annealing process to find minima in the total free-energy. The parameters are sampled with a uniform distribution with the following limits. 

\begin{align}
    A &\in[-0.1,0.1]\\
    \mu &\in[-1,1]\\
    \sigma^2 &\in[0.1,2]\\
    B &\in[-0.1,0.1]\\
    \rho &\in[-1,1]\\
    \Sigma^2 &\in[0.1,2]
\end{align}

\section{Section 4: Discussion on boundary conditions and topological constraints.}

In this work, we studied patches with rotational symmetry, which enforce the existence of a charge $+1$ topological defect at their core. This is to reflect topological defects that have been associated with morphological changes in experimental systems~\cite{keber2014topology,endresen2021topological,maroudas2021topological,blanch2021quantifying,ravichandran2024topology}. Here, we address the question of how these patches can be tiled together to represent a closed surface, such as those often observed in naturally occurring biological systems.


The total topological charge of a vector field on a closed surface is constrained according to the Poincar\'{e}-Hopf Theorem, which states that the sum of the topological charges of a vector field is equal to the Euler characteristic of the surface on which they lie. Therefore a simple surface with a spherical topology can be tiled with two $+1$ defects as the Euler characteristic of a spherical surface is $+2$. More generally, if we want to tile a spherical surface with many patches containing $+1$ topological defects, additional negative charges are required to meet the constraints of the Poincar\'{e}-Hopf Theorem. This is observed in regenerating freshwater Hydra, on which morphogenesis appears to be driven by $+1$ defects which are placed according to chemical morphogens, and the remaining $-1/2$ defects arise as a result of the Poincar\'{e}-Hopf Theorem and the nematic nature of the interactions between the filaments~\cite{maroudas2021topological,ravichandran2024topology}. 

Interestingly, a similar statement can be made about the Gaussian curvature, the integral of which is also proportional to the Euler characteristic of the surface (for a closed surface) via the Gauss-Bonnet Theorem. Thus the total topological charge on a surface and the total Gaussian curvature are equal~\cite{bowick2009two}. Therefore topological defects on closed surfaces merely redistribute the Gaussian curvature that is already present on the surface. In this scenario, the additional negative defects necessitated by the Poincar\'{e}-Hopf Theorem can simply be placed in the negative Gaussian curvature regions of the surface necessitated by the Gauss-Bonnet Theorem; which reduces their internal energy. This is again consistent with observations on freshwater Hydra, on which the $-1/2$ defects occur in the regions of maximally negative Gaussian curvature~\cite{maroudas2021topological,ravichandran2024topology,pearce2020defect}. 

\section{Section 5: Parameters and boundary conditions for all numerical results}

All simulations are performed on a discretized line of $N=100$ points. Parameter values and boundary conditions for all numerical work is provided in the tables below. 

\begin{center}
\begin{tabular}{ |c|c|c| } 
\hline
Figure & Parameters & Constraints \\
\hline
\multirow{2}{4em}{Fig. 1b} & $k_1=k_2=k_3=k/3$ & $\zeta''(R) = 0$\\ 
& $\sigma+k+k_B = 1$ & $\psi''(R) = 0$\\ 
\hline
\hline
\multirow{3}{4em}{Fig. 1c} & $k_1+k_3=1$ & $\zeta''(R) = 0$\\ 
& $\sigma=k_B = 1/10$ & $\psi'(r) = 0$\\ 
& $k_2=0$ & \\ 
\hline
\hline
\multirow{3}{4em}{Fig. 1d} & $k_1=k_3=1/2$ & $\zeta''(R) = 0$\\ 
& $\sigma=k_B = 1/10$ & $\psi(r) = 0$\\ 
& $k_2=0$ & $\psi'(r) = 0$\\ 
\hline
\hline
\multirow{3}{4em}{Fig. 1e} & $k_3=1$ & $\zeta''(R) = 0$\\ 
& $\sigma=k_B = 1/10$ & $\psi(r) = \pi/8$\\ 
& $k_1=k_2=0$ & $\psi'(r) = 0$\\ 
\hline
\end{tabular}
\end{center}

\begin{center}
\begin{tabular}{ |c|c|c| } 
\hline
Figure & Parameters & Constraints \\
\hline
\multirow{2}{4em}{Fig. 2a Fig. 2b} & $k_1+k_2+k_3=1$ & $\zeta''(R) = 0$\\ 
& $\sigma=k_B = 1/10$ & $\psi''(R) = 0$\\ 
\hline
\hline
\multirow{4}{4em}{Fig. 2c} & $k_1=1/2$ & $\zeta''(R) = 0$\\ 
& $k_2=1/10$ & $\psi''(R) = 0$\\ 
& $k_3=4/10$ & \\ 
& $\sigma=k_B = 1/10$ & \\ 
\hline
\hline
\multirow{4}{4em}{Fig. 2d} & $k_1=13/20$ & $\zeta''(R) = 0$\\ 
& $k_2=1/10$ & $\psi''(R) = 0$\\ 
& $k_3=1/4$ & \\ 
& $\sigma=k_B = 1/10$ & \\ 
\hline
\end{tabular}
\end{center}

\begin{center}
\begin{tabular}{ |c|c|c| } 
\hline
Figure & Parameters & Constraints \\
\hline
\multirow{2}{4em}{Fig. 3a Fig. 3b} & $k_1+k_2+k_3=1$ & $\zeta'(R) = 0$\\ 
& $k_2=1/3$ & $\psi(R) = \psi_R$\\ 
& $\sigma=k_B = 1/10$ & $\psi'(R) = 0$\\ 
\hline
\hline
\multirow{4}{4em}{Fig. 3c Fig. 4b} & $k_1=1/2$ & $\zeta'(R) = 0$\\ 
& $k_2=1/3$ & $\psi(R) = 0$\\ 
& $k_3=1/6$ & $\psi'(R) = 0$\\ 
& $\sigma=k_B = 1/10$ & \\ 
\hline
\hline
\multirow{4}{4em}{Fig. 3d Fig. 4c} & $k_1=2/3$ & $\zeta'(R) = 0$\\ 
& $k_2=1/3$ & $\psi(R) = 0$\\ 
& $k_3=0$ & $\psi'(R) = 0$\\ 
& $\sigma=k_B = 1/10$ & \\ 
\hline
\hline
\multirow{4}{4em}{Fig. 3e} & $k_1=0$ & $\zeta'(R) = 0$\\ 
& $k_2=1/3$ & $\psi(R) = \pi/2$\\ 
& $k_3=2/3$ & $\psi'(R) = 0$\\ 
& $\sigma=k_B = 1/10$ & \\ 
\hline
\end{tabular}
\end{center}

\begin{center}
\begin{tabular}{ |c|c|c| } 
\hline
Figure & Parameters & Constraints \\
\hline
\multirow{2}{4em}{Fig. 4a Black} & $k_1=1/2$ & $\zeta'(R) = 0$\\ 
& $k_3=1/6$ & $\psi(R) = 0$\\ 
& $\sigma=k_B = 1/10$ & $\psi'(R) = 0$\\ 
\hline
\hline
\multirow{2}{4em}{Fig. 4a Cyan} & $k_1=2/3$ & $\zeta'(R) = 0$\\ 
& $k_3=0$ & $\psi(R) = 0$\\ 
& $\sigma=k_B = 1/10$ & $\psi'(R) = 0$\\ 
\hline
\hline
\multirow{2}{4em}{Fig. 4a Magenta} & $k_1=0$ & $\zeta'(R) = 0$\\ 
& $k_3=2/3$ & $\psi(R) = \pi/2$\\ 
& $\sigma=k_B = 1/10$ & $\psi'(R) = 0$\\ 
\hline
\end{tabular}
\end{center}

\begin{center}
\begin{tabular}{ |c|c|c| } 
\hline
Figure & Parameters & Constraints \\
\hline
\multirow{4}{4em}{Fig. 4d} & $k_1=1/2$ & $\zeta'(R) = 1$\\ 
& $k_2=1/3$ & $\psi(R) = 0$\\ 
& $k_3=1/6$ & $\psi'(R) = 0$\\ 
& $\sigma=k_B = 1/10$ & \\ 
\hline
\hline
\multirow{4}{4em}{Fig. 4e} & $k_1=2/3$ & $\zeta'(R) = 1$\\ 
& $k_2=1/3$ & $\psi(R) = 0$\\ 
& $k_3=0$ & $\psi'(R) = 0$\\ 
& $\sigma=k_B = 1/10$ & \\ 
\hline
\hline
\multirow{4}{4em}{Fig. 4f} & $k_1=0$ & $\zeta'(R) = 1$\\ 
& $k_2=1/3$ & $\psi(R) = \pi/2$\\ 
& $k_3=2/3$ & $\psi'(R) = 0$\\ 
& $\sigma=k_B = 1/10$ & \\ 
\hline
\end{tabular}
\end{center}

\begin{center}
\begin{tabular}{ |c|c| } 
\hline
Figure & Parameters \\
\hline
\multirow{4}{4em}{Fig. S1a} & $k_1=k_2=k_3=k/3$ \\ 
& $k+\sigma+k_B=1$ \\ 
& $\psi=0$ \\ 
& $\Delta=1/100$ \\ 
\hline
\hline
\multirow{4}{4em}{Fig. S1b} & $k_1+k_3=1$ \\ 
& $k_2=0$ \\
& $\sigma=k_B=1/10$ \\ 
& $\Delta=1/100$ \\ 
\hline
\hline
\multirow{4}{4em}{Fig. S1c} & $k_1+k_2+k_3=1$ \\ 
& $\sigma=k_B=1/10$ \\ 
& $\psi=0$ \\ 
& $\Delta=1/100$ \\ 
\hline
\end{tabular}
\end{center}

\section{Section 6: Quantitative comparison to biological systems}

In this section, we summarise the main arguments to make a comparison with biological systems. The values of the material and geometrical parameters can be found in Table~\ref{tab:table1}. Due to the complexity of the systems considered and the simplifications required for our theoretical treatment, we consider only orders of magnitude.

Several biological systems share the main characteristics of a fluid membrane with nematic order. In particular, a quasi-two dimensional geometry, a fluid-like behaviour in the experimental time scale, and nematic orientational order over macroscopic time and length scales. In the following, we focus on two special cases: films of cytoskeletal filaments and thin tissues.

First, let us consider a thin layer made of cytoskeletal filaments, like actin or microtubules, on top of a supported lipid bilayer, \cite{sanchez2012spontaneous,sciortino2021pattern,memarian2021active}. For a lipid bilayer, the bending rigidity ranges from $k_B=10^{-8}-10^{-7}$~$\mu N \mu m$, and the surface tension ranges from $\sigma=10^{-6}-10^{-3}~N/m$ \cite{evans1987physical,rawicz2000effect,roux2010membrane,phillips2009emerging}. Actin and microtubule films exhibit nematic phases \cite{sanchez2012spontaneous,sciortino2021pattern,memarian2021active}, and the Frank constant $K$ for both filament types has been measured. Specifically, a thin film of actin filaments on an oil-water interface has a Frank constant ranging from $K=10^{-1}-1~pN$, \cite{zhang2018interplay,yadav2019filament}. The reduced Frank constant $k=K h$ depends also on the thickness of the layer $h$. Taking $h=0.1~\mu m$, which is typical thickness of an actin cortex in cells \cite{hanakam1996myristoylated,clark2013monitoring,laplaud2021pinching}, leads to $k=10^{-8}-10^{-7}~\mu N \mu m$. For thin films of microtubules on an oil-water interface, the reduced Frank constant was $k=10^{-3}~\mu N \mu m$ and the layer thickness was $h=10~\mu m$ \cite{velez2023probing}. The corresponding Frank constant for the latter case is compatible with that found in spindles of Xenopus egg \cite{oriola2020active}. 

Next, let us consider a monolayer (or thin multilayer) of cells that can be supported by a thin elastic substrate, \cite{harris2012characterizing,maroudas2021topological,da2022fibroblasts}. For epithelial cells, the bending rigidity is measured as $k_B=10^{-1}-1$~$\mu N \mu m$, and the tension can vary from $\sigma=10^{-3}-10^{-1}~N/m$ \cite{harris2012characterizing,latorre2018active,trushko2020buckling,fouchard2020curling,duque2023fracture}. Several types of cells, including epithelial or fibroblast, can exhibit nematic phases, \cite{duclos2014perfect,saw2017topological,armengol2023epithelia}. At time of writing, neither the Frank constant nor the reduced Frank constant has been directly measured in tissues, but several works estimated their values from indirect experimental measurements. The value of the Frank constant depends on the cell type, being on the range of $K=10^{3}-10^4~pN$ for epithelial cells \cite{lee2011crawling,perez2019active} and on the scale of $K=10^5~pN$ for fibroblast cells \cite{duclos2017topological,duclos2018spontaneous}. Taking $h=10~\mu m$, which is the typical thickness of a cell monolayer, leads to the values of the reduce Frank constant on the range of $k=10^{-2}-10^{-1}~\mu N \mu m$ for epithelial cells and on the scale of $K=1~\mu N \mu m$ for fibroblast cells.

\begin{table*}[b]
\begin{ruledtabular}
\begin{tabular}{|c|cccc|}
& Surface tension & Bending rigidity &Length Scale& Reference \\ & $\sigma[N/m]$ & $k_B[\mu N \mu m]$ & $\sqrt{k_B/\sigma}[\mu m]$& \\ \hline
Lipid vesicles & $10^{-6}-10^{-3}$& $10^{-8}-10^{-7}$ & $10^{-2}-10^{-1}~^{(*)}$ & \cite{evans1987physical,rawicz2000effect,roux2010membrane,phillips2009emerging} \\
\hline
Epithelial cell monolayer & $10^{-3}-10^{-1}$ & $10^{-1}-1$ & $1-10~^{(*)}$  & \cite{harris2012characterizing,latorre2018active,trushko2020buckling,fouchard2020curling,duque2023fracture} \\
\hline
\hline
& Frank constant & Thickness & Reduced Frank constant & Reference \\ & $K[pN]$ & $h[\mu m]$ & $k[\mu N \mu m]$ & \\ \hline
F-actin film  & $10^{-1}-1$& $10^{-1}$ & $10^{-8}-10^{-7} ~^{(*)}$ &\cite{zhang2018interplay,yadav2019filament} \\
at the oil-water interface &  & & & \\
\hline
Microtubule film  & $10^{2} ~^{(*)}$& $10$ & $10^{-3}$ & \cite{guillamat2016probing,velez2023probing}\\
at the oil-water interface & & && \\
\hline
Xenopus egg
  & $10^{2}$& - & - & \cite{oriola2020active}\\
extract spindles & & && \\
\hline
Epithelial cell monolayer & $10^3-10^4~^{(*)}$ & $10$ & $10^{-2}-10^{-1}~^{(*)}$ &\cite{lee2011crawling,perez2019active} \\
\hline
Fibroblast cell monolayer & $10^5~^{(*)}$ & $10$ & $1~^{(*)}$ &\cite{duclos2017topological,duclos2018spontaneous} \\
\end{tabular}
\end{ruledtabular}
\caption{Table of material parameters and geometrical parameters of some biological systems. The symbol $*$ denotes the values that were estimated. For a thin layer, the reduced Frank constant $k=K*h$ depends on the Frank constant $K$ and the thickness of the layer $h$.}
\label{tab:table1}	
\end{table*}

We show that the instability threshold of a flat disc with an aster at its center is found to depend on the elastic parameters of the fluid membrane and embedded nematic field, in particular the surface tension $\sigma$, the bending rigidity $k_B$, and the reduced Frank constant $k$, as well as, geomtrical parameters, such as the radius of the disc $R$. For simplicity, we consider that all the reduced Frank constants are of the same order of magnitude, thus we take the one-constant approximation (i.e. $k_1=k_2=k_3\sim k$). Next, we evaluate this threshold for the previous biological systems in two limiting cases controlled by the ratio $k_B/\sigma R^2$.

In the bending dominated regime, $k_B\gg \sigma R^2$, the instability threshold is dependent on the ratio $k/k_B$ and the critical magnitude is of order of $(k/k_B)_c\sim1$, see Eq.~(4) in the main text. For a film of actin filaments, we expect that this ratio is $k/k_B=1$. If the film is made of microtubules, then the ratio ranges from $k/k_B=10^4-10^5$, which is above the instability threshold. In microtubule films the thickness of the layer can be modulated and may allow to control the value of this ratio $k/k_B$ \cite{guillamat2016probing,velez2023probing}. For tissues and ignoring the elasticity of the underlying substrate, the ratio is expected to be $k/k_B=10^{-1}-1$.

In the tension dominated regime, $k_B\ll \sigma R^2$, the instability threshold of a flat interface is controlled by the ratio $k/\sigma R^2$, and the critical value is of order $(k/\sigma R^2)_c \sim 1$. One can estimate an upper bound by replacing the radius $R$ with the layer thickness $h$, which is the smallest lengthscale in the problem. Depending on the surface tension $\sigma$, the upper bound is $k/\sigma h^2 = 10^{-3}-10$ for a film of actin filaments and $k/\sigma h^2 = 10^{-2}-10$ for a film of microtubules.  Ignoring the elasticity of the underlying substrate, the upper bound for a cell monolayer is $k/\sigma h^2 = 10^{-3}-10$. In all cases, the upper bound is at most a factor $10$ larger than the critical value, which suggest that in low-tension regimes, these cases can be close to the instability threshold.

Note that because the geometry is quasi-two dimensional, the thickness of the layer $h$ needs to be smaller than the typical radius $R$. Therefore we can estimate an upper bound for the ratio $k_B/\sigma R^2$. For both biological examples, we found that $k_B/\sigma h^2\leq 1$, suggesting that experimental cases are in a tension dominated regime $k_B/\sigma R^2\ll 1$.

To summarise, in this section we evaluated the proximity to the instability threshold of a flat fluid membrane with an aster topological defect for two biological systems: films of cytoskeletal filaments on a supported lipid bilayer and cell monolayers on a supported elastic substrate. In the bending dominated regime $k_B\gg \sigma R^2$, both films of actin filaments and cell monolayers can be close to the instability threshold, and films of microtubules are above the instability threshold. In the tension dominated regime  $k_B\ll \sigma R^2$, we found that all these biological examples can be close to the instability threshold. Our arguments here focus on orders of magnitude, to know whether a flat membrane can be rendered unstable for the cases that are close to the instability threshold would depend on the precise values of the elastic and geometrical parameters. Two possible ways to cross the instability threshold in these cases is to either operate in the bending dominated regime, by for instance reducing the surface tension of the supporting substrates, or to increase the reduced Frank constant by for instance increasing the number density of cytoskeletal filaments or cells.

\end{document}